\documentclass[
aps,%
8pt,%
final,%
notitlepage,%
oneside,%
onecolumn,,%
nobibnotes,%
nofootinbib,% 
superscriptaddress,%
noshowpacs,%
centertags]%
{revtex4}
\usepackage{amssymb}
\usepackage{multirow}
\usepackage{natbib}
\usepackage{graphicx}
\usepackage{longtable} 

\begin{document}
\selectlanguage{english}

\title{Non-LTE Sodium Abundance in Galactic Thick- and Thin-Disk Red
Giants}% Разбиение на строки осуществляется командой \\

\author{\firstname{S.~A.}~\surname{Alexeeva}}
\email{alexeeva@inasan.ru}
\affiliation{Institute of Astronomy RAS, Moscow, Russia}
%\affiliation{Institute of Astronomy RAS, Moscow, Russia}
\author{\firstname{Yu.~V.}~\surname{Pakhomov}}
\email{pakhomov@inasan.ru}
\affiliation{Institute of Astronomy RAS, Moscow, Russia}
%\affiliation{Institute of Astronomy RAS, Moscow, Russia}
\author{\firstname{L.~I.}~\surname{Mashonkina}}
\email{lima@inasan.ru}
\affiliation{Institute of Astronomy RAS, Moscow, Russia}

%\date{\today}
%\today печатает cегодняшнее число

\begin{abstract}
We evaluate non-local thermodynamical equilibrium (non-LTE) line formation for neutral sodium in model
atmospheres of the 79 red giants using the model atom that incorporates the best available atomic data.
The non-LTE abundances of Na were determined from Na I 6154, 6161\,\AA\, for the 38 stars of the thin disk 
(15 of them the BaII stars), 15 stars of the thick disk, 13 stars of Hercules stream
and 13 transition stars which can be identified with neither thin disk nor thick disk.
For Na I 6154, 6161\,\AA\, non-LTE abundance corrections amount to -0.06 to -0.24 dex 
depending on stellar parameters. 
We found no difference in [Na/Fe] abundance between the thick disk and thin disk and the obtained abundances are close to the solar one. We confirmed a weak excess of [Na/Fe] in BaII stars. Stars of the Hercules 
stream reveal [Na/Fe] abundances close to the solar one. The obtained results can be used to constrain the nucleosynthesis models for Na. 

\end{abstract}

\maketitle

\section{INTRODUCTION}

Studies of stellar sodium abundances play an important role 
in better understanding of various fields of astrophysics like nucleosynthesis, stellar 
evolution and galacto-chemical evolution. 

To solve the nucleosynthesis problems, it is important to understand in which thermonuclear reactions
sodium is synthesized. 
Sodium can be synthesized in the carbon-burning reactions  ($^{12}$C+$^{12}$C$\to$ $^{23}$Na+p) into the interior of massive stars. In this case,
the production rate does not depend on the abundance of metals in the previous generation of stars
(Woosley and Weaver 1995) and sodium belongs to the primary elements. On the other hand, sodium can be synthesized in the NeNa cycle ($^{22}$Ne+p$\to$ $^{23}$Na+$\gamma$) in all stars with $М$~>~1.5~$M_\odot$ or in the
$\alpha$-reaction ($^{14}$N+$\alpha$ $\to$ $^{19}$F+$\gamma$, $^{19}$F+$\alpha$ $\to$ $^{23}$Na+$\gamma$) in asymptotic-giant-branch (AGB) stars. In the NeNa cycle, the production rate depends on
the excess of protons that is determined by the initial abundance of metals (Denisenkov and Denisenkova 1990). In these cases, sodium is considered as a secondary element.
Probably, the first type of reactions does not contribute to the production of Na because an underabundance Na
compared with iron is observed in the metal-poor stars (Cayrel et al. 2004; Andrievsky
et al. 2007).
Determination and analysis of changes in sodium abundances over time should 
clarify the types of thermonuclear reactions, mechanisms and Na production rates.

Investigating the differences in sodium abundance between the thin and thick disks is important for
understanding the chemical evolution of the Galaxy. As a rule, main-sequence dwarf stars are used for this
purpose, because the sodium abundance in dwarfs reflects the value that was in the protostellar cloud
as a result of the chemical evolution of the previous generation of stars. The goal of this paper is to
show that low-mass giants ($М$~<~2.0~$M_\odot$) can also be used to solve the problems of Galactic chemical
evolution, because the NeNa-cycle reactions in them are inefficient. It should be noted that giants have
their advantage in luminosity, which is higher than that for dwarfs. Consequently, invoking giants would
allow the Galaxy to be scanned at longer distances.

According to the studies of dwarf stars with [Fe/H]>$-$1 by
Reddy et al. (2003) and Bensby et al. (2003), no difference in [Na/Fe] between the thin
and thick disks is detected.
Alves-Brito et al. (2010) detected no [Na/Fe] differences between thick- and thin-disk red giants either.
At the same time, when investigating red giants based on the LTE approach, Pakhomov (1012, 2013) showed that the [Na/Fe]
abundance in the thin disk is higher than that in the thick one. The author concludes that this is a
consequence of the difference in the abundance of neon from which sodium is formed. In this paper, we show that this is not the only possible cause of this difference.

One of the subjects of stellar evolution research is the formation of a vast convective envelope extending
to the stellar core at the red-giant stage, which causes the nuclear reaction products to be brought to the
surface. Theoretical studies of deep mixing were performed by Denissenkov and Weiss (1996), Denissenkov and Herwig (2003), and Charbonnel and Lagarde (2010). They showed that the sodium produced in the NeNa cycle in hydrogen-burning regions could
be mixed with the stellar surface layers during the first dredge-up (1DUP). In this case, an enhanced [Na/Fe] abundance can serve as an indicator that this stage of stellar evolution has been reached. We attempted to detect [Na/Fe] differences in thin-disk giants of various masses.

Undoubtedly,  it seems interesting to compare the [Na/Fe] abundances between dwarfs
and giants. In this sense, our paper is the ground work for future studies of dwarfs with the goal of comparison with giants by a unified method. Some of the studies showed [Na/Fe] in the atmospheres of
red giants to be higher than that in the atmospheres of dwarfs (Boyarchuk et al. 2001; Pasquini et al. 2004).
However, giants have more rarefied atmospheres than dwarfs and, hence, non-LTE effects will manifest
themselves differently in the atmospheres of these stars. Thus, there is a need to invoke the non-LTE
approach for comparing dwarf stars and giant stars between themselves. Likewise, the metallicities of
thick-disk stars are also, on average, lower than those of thin-disk stars, although they have an overlapping
region (Fuhrmann 1998). Since non-LTE effects can manifest themselves differently in stars of different metallicities, we abandon using this assumption when determining the sodium abundance in Galactic
thin- and thick-disk stars and apply the so-called non-LTE approach (Mihalas 1978).

The original methods of non-LTE calculations for Na~I were developed by Gehren (1975), Boyarchuk et al. (1985), Mashonkina et al. (1993), Baumuller et al. (1998), Gratton et al. (1999), Takeda et al. (2003), Shi et al. (2004), and Lind et al. (2011). Unfortunately, the results obtained by different authors do not agree between themselves. For example,
Gratton et al. (1999) obtained mostly positive corrections for giant stars. According to these authors,
the LTE abundances turn out to be underestimated by about 0.3 dex. 
Mashonkina et al. (2000) and other authors obtained negative non-LTE corrections, from $-$0.05 dex for dwarf stars
to $-$0.20 dex for giant stars. Thus, those who applied the corrections from
Gratton et al. (1999) derived high sodium abundances
with [Na/Fe]$_{non-LTE}$=$+0.37$ (Carretta et al. 2003) and [Na/Fe]$_{non-LTE}$=$+0.48$ (Gratton et al. 2012).
Accordingly, those who applied the corrections from Mashonkina et al. (2000) derived nearly solar sodium 
abundances with [Na/Fe]$_{non-LTE}$=$+0.07$ (Munoz et al. 2013). For this reason, the results of different
authors are difficult to compare. In addition, there are researchers who prefer to refrain from applying the non-LTE approach, for example, Johnson and Pilachowski (2012).
 
In this paper we improved the Na~I model atom by using new, more accurate atomic data for collisional
processes with electrons and hydrogen atoms. By applying the non-LTE approach, we determined the Na
abundance for stars from various kinematic groups of our Galaxy (the thin disk, the thick disk, the Hercules
stream) and stars that cannot be attributed neither to the thick disk nor to the thin one. We will call these stars the transitional ones for short.
 
Among the 38 investigated thin-disk stars, 15 stars are barium ones; therefore, they were isolated into a separate subgroup by their chemical properties. Ba~II stars are stars with chemical peculiarities that
have attracted the attention of researchers for more than 60 years (Bidelman and Keenan 1951). As a rule,
Ba~II stars are detected only in the thin disk. One of the hypotheses for the formation of Ba~II stars is
related to the binarity of these objects. Being on the AGB, the more massive component produces chemical elements in the $s$-process, which are transferred to the secondary companion, a main-sequence star, when the envelope is ejected. At the final stage of its evolution, the primary becomes a hard-to-detect white dwarf, while the secondary evolves into a red
giant. The name of a Ba~II star is associated with enhanced Ba lines in its spectrum. This hypothesis is
supported by the observations confirming the binarity of Ba~II stars (McClure and Woodsworth 1990),
but it is inconsistent with the high eccentricities of their orbits. This is indicative of our incomplete understanding of how the Ba~II stars are formed. However, it is worth noting that a mechanism for the formation of high orbital eccentricities of Ba~II stars associated with the birth of a white dwarf has been proposed not so long ago (Davis et al. 2008). Most of the works on the chemical composition of Ba~II stars are based on the LTE method, with the emphasis being on heavy elements. The studies of sodium in Ba~II stars are few. Differing results were obtained in these few works: Antipova et al. (2003, 2005) and Pereira et al. (2011) found a [Na/Fe] overabundance; Liu et al. (2009) found a solar [Na/Fe] value. Clayton (2003) explained the overabundance of sodium by its synthesis in the $s$-process. On the other hand, the overabundance can be due to non-LTE effects. Determining the non-LTE sodium abundance in Ba~II
stars is of interest from the viewpoint of testing the assumption of sodium production at the AGB stage and the possibility of its transfer in a binary system.

Our sample also includes 13 Hercules-stream stars. The Hercules stream is a group of stars moving with a velocity that is shifted in U component from the main velocity distribution in the solar neighborhood (Dehnen 1998; Famaey et al. 2005). In its kinematic properties, the Hercules stream lies between the thick and thin disks. The nature of this moving stream has
not yet been clarified. As a rule, the stellar streams are remnants of open star clusters. The stream stars have similar ages and elemental abundances. The second hypothesis about the origin is related to the resonant interactions of the Galactic bar. This is the so-called dynamical nature of the Hercules stream (Fux 2001), whose stars are not associated by
common evolution and originate from the inner-disk regions. Another hypothesis is based on the fact that the Hercules stream was formed when our Galaxy merged with another system. The chemical composition of Hercules-stream stars was studied by the LTE method for dwarf stars by Soubiran and Girard (2005) and for red giants by Pakhomov et al. (2011); these
studies showed a significant difference between the ages and elemental abundances of the stream stars.
The goal of our work is also to redetermine the non-LTE sodium abundance in the giants investigated by Pakhomov et al. (2011). This will allow the Hercules-stream stars to be qualitatively compared with stars of various populations in our Galaxy for the sodium abundance in them, which will possibly be useful for
unraveling their nature. 

The paper is organized as follows. In the next section, we consider an improved Na~I model atom,
discuss the formation of lines under non-LTE, and test the model atom using the Sun as a standard star. 
We present the sodium abundance determinations in stars. We discuss the results for thin- and thick-
disk stars, Ba~II stars, Hercules-stream stars, and transitional stars and compare them with theoretical
predictions and other studies.

\section{non-LTE LINE FORMATION FOR Na~I}

\subsection{The Model Atom and Atomic Data}

{\bf Model atom.} 
The model atom consists of 17 Na~I energy levels up to $n$ = 7, $l$ = 5 and the ground
state of Na~II. The level energies were taken from Martin and Zalubas (1981). The uppermost levels is
below the ionization threshold by 0.29 eV. All states with $n$ = 7 have close energies and their populations
must be equilibrium ones relative to one another; therefore, they were combined into one superlevel. The fine structure was neglected, except for the 3p$^{2}$P$^{o}$ state. Our model atom includes 68 allowed bound-bound transitions, given the transitions to the levels included in the combined ones.

{\bf Radiative transitions.} For 68 allowed bound-bound transitions
adopting oscillator strengths were taken from calculations of C. Froese Fischer\footnote[1]{
http:$//$www.vuse.vanderbilt.edu$/$}. The photoionization cross sections for the levels with $l \le 4$ were taken from the
TOPbase database (Cunto and Mendoza 1992)\footnote[2]{http:$//$legacy.gsfc.nasa.gov$/$topbase}.
For the high excited combined levels, we adopted the hydrogen-like cross sections.

{\bf Collisions with electrons.} 
For excitation and ionization of Na~I by electron collisions 
we adopted cross-sections, calculated with new convergent close-coupling approach (CCC) by Igenbergs et al. (2008) including real spectroscopic states 3s, 3p, 4s, 3d, 4p, 5s, 4d, 4f. For transitions from the levels above 5s we adopted 
empirically adjusted rates by Park (1971). 

{\bf Collisions with hydrogen.} 
Neutral hydrogen is the most abundant type of atoms in the atmospheres of cool stars. Inelastic collisions
with hydrogen atoms are less efficient than electron collisions. However, their role may be great because
of the large number of hydrogen atoms. We took into account the inelastic collisions with neutral
hydrogen atoms using the rate coefficients from Barklem et al. (2010) for all of the possible transitions between the nine lowest levels, namely 3s, 3p, 4s, 3d, 4p, 5s, 4d, 4f, and 5p, and the charge exchange
processes Na~I+H~I(1s) $\rightleftarrows$ Na$^+$+H$^-$. Lind et al. (2011) showed that the rate coefficients from
Barklem et al. (2010) are smaller than those derived from the Drawin (1968) formula by several orders of magnitude (from 1 to 6, depending on the transition).

\subsection{Mechanisms of Departures from LTE for Na~I}
 
For solving combined radiative transfer and statistical equilibrium equations
we used the DETAIL program of Butler $\&$ Giddings (1985) based on the accelerated $\Lambda$ -- iteration (ALI)
scheme following the efficient method described by Rybicki $\&$ Hummer (1991, 1992).
The opacity package of the DETAIL code was recently updated as outlined by Mashonkina et al. (2011).
Synthetic line profiles were computed via the SIU (Spectrum Investigation Utility) program (Reetz, 1991) with invoking the departure coefficients, b$_i$ = n$_{non-LTE}$/n$_{LTE}$, where n$_{non-LTE}$ and n$_{LTE}$ are the statistical equilibrium and
thermal equilibrium number densities, respectively. 

\begin{figure}[h]
\begin{center}
\parbox{0.8\linewidth}{\includegraphics[scale=0.7]{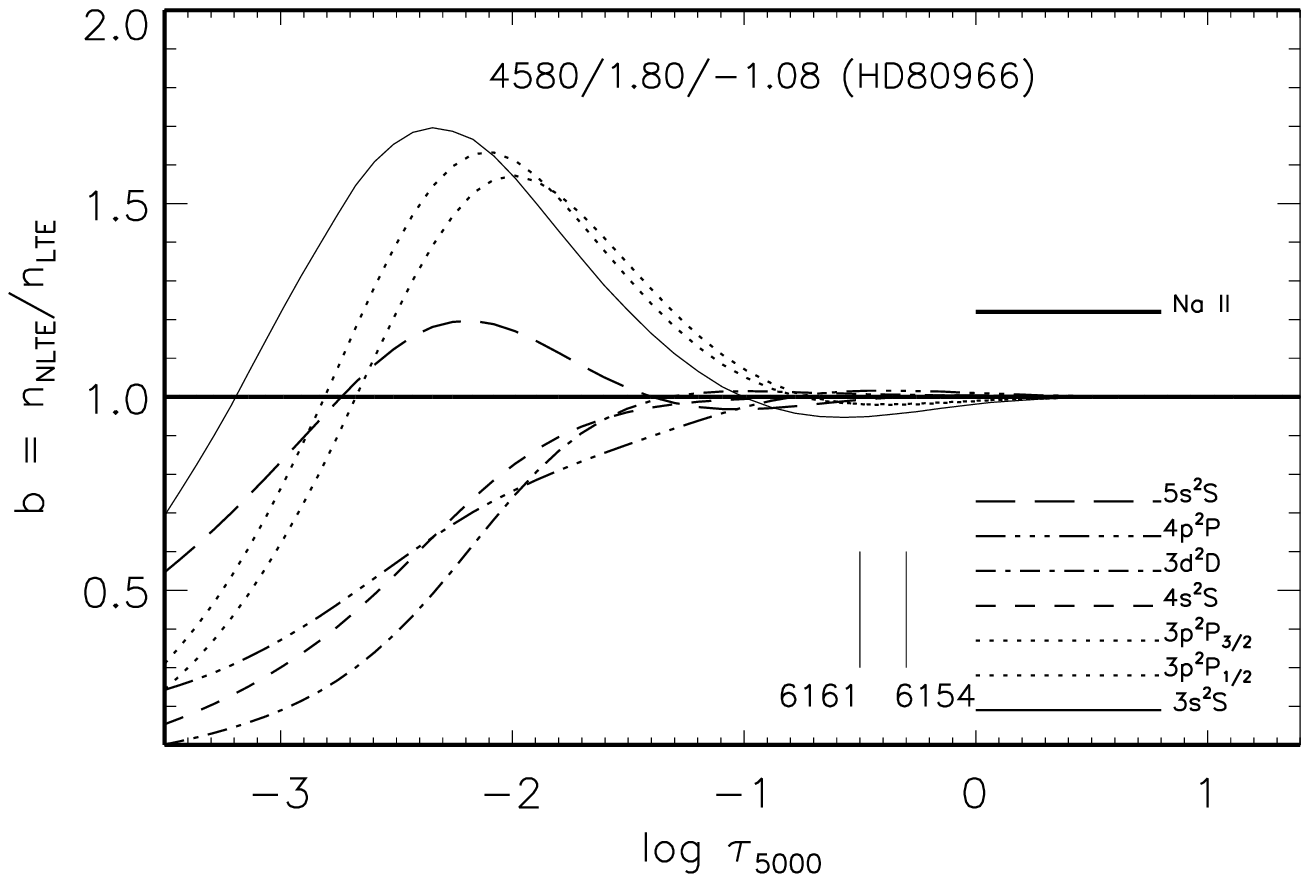}\\
\centering}
\hfill
%\\[0ex]
\parbox{0.8\linewidth}{\includegraphics[scale=0.7]{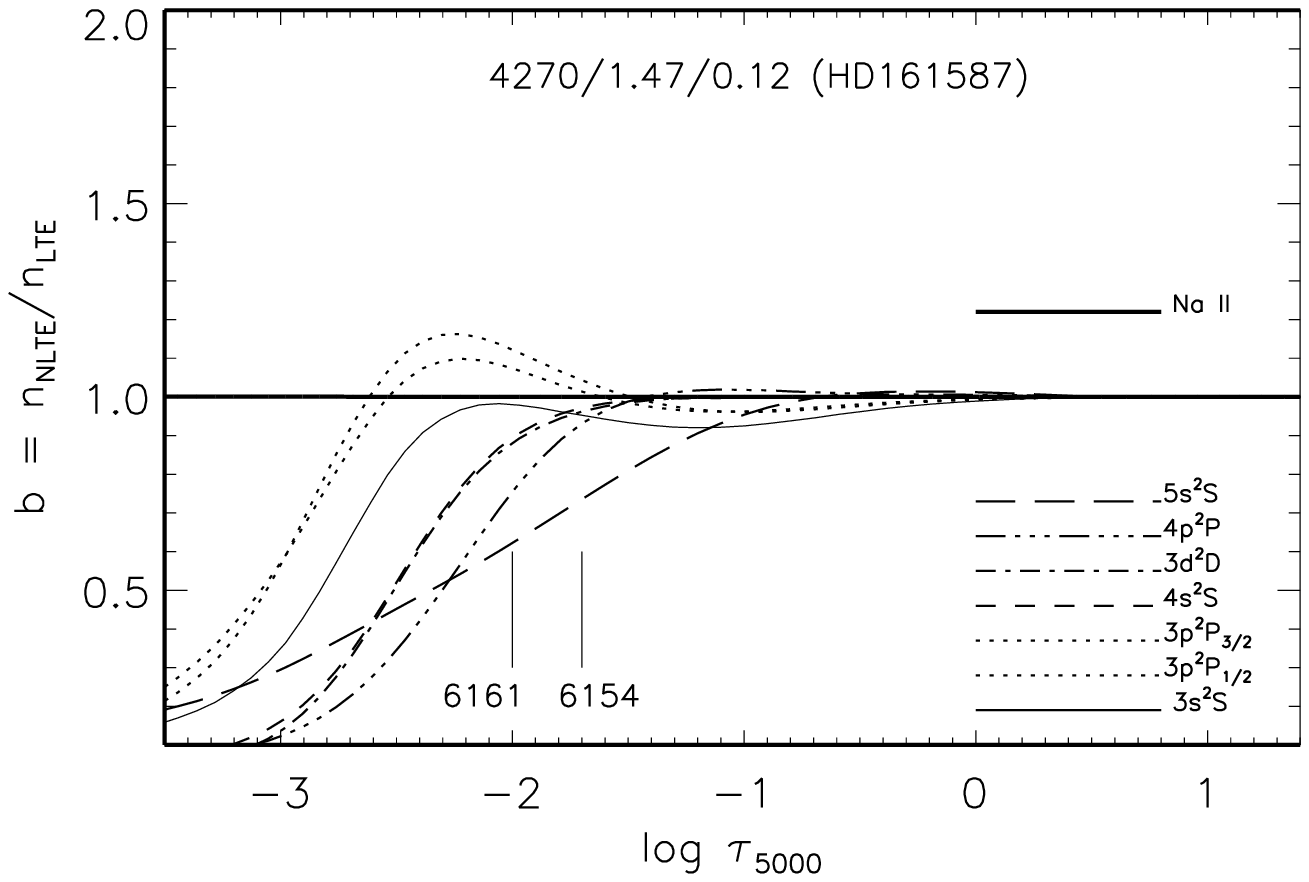}\\
\centering}
\hfill
\end{center}
\caption{ Departure coefficients for the Na~I levels versus optical depth $\tau_{5000}$ 
in the model atmospheres with $T_{\rm{eff}}$=4580K, log$g$=1.80, [Fe/H]=$-$1.08 (HD 80966, panel (a)) and
 $T_{\rm{eff}}$=4270K, log$g$=1.47, [Fe/H]=$+$0.12 (HD 161587, panel (b)). The vertical lines mark the depths of line core 
 formation, i.e., the depths where $\tau$ = 1 at the central wavelengths.}
\label{b}
\end{figure}

Figure \ref{b} presents the departure coefficients for the Na~I levels in the atmospheres of two investigated stars, HD 80966 ($T_{\rm{eff}}$=4580~K, log$\, \textsl{g}$=1.80, [Fe/H]=$-$1.08) and HD~161587 ($T_{\rm{eff}}$=4270~K, log$\, \textsl{g}$=1.47, [Fe/H]=$+$0.12). Bruls et al. (1992) showed that various mechanisms compete in the establishment of
statistical equilibrium for any atom: the recombinations to excited states and the subsequent cascade transitions to lower levels, the escape of line photons at depths that do not exceed the mean free path, the absorption of photons at depths where the medium is opaque in the line core but is already transparent
in the wings, and excessive photoionization at depths where $\tau$< 1 behind the level ionization threshold.

We investigate the 6154 and 6161 \,\AA\, lines formed at the 3p$^{2}$P$^{o}$$_{1/2}$ -- 5s$^{2}$S and 3p$^{2}$P$^{o}$$_{3/2}$ -- 5s$^{2}$S transitions.
Analyzing the departure coefficients of these levels (Fig. \ref{b}) for HD 161587, we see that the 
departure coefficient of the lower level $b_{l}$ is larger than the departure coefficient
of the upper level $b_{u}$ and that the inequality $b_{l}~>~1$ holds for the lines being investigated in their formation regions. Both effects cause the absorption line in non-LTE to be stronger than that in LTE. This means that the non-LTE abundance corrections ($\Delta_{non-LTE}$=log$\epsilon_{LTE}$--log$\epsilon_{non-LTE}$) will be negative for the Na~I 6154 and
6161 \,\AA\, lines. In contrast, in the atmosphere of HD 80966, the formation regions of the lines being investigated are in layers where, on the one hand, there is excessive photoionization from the 3p level ($\lambda$ $_{\rm{thr}}$ = 4086 \,\AA\,) and, on the other hand, the 3s and 3p levels become overpopulated due to the recombinations to highly excited states followed by the cascade transitions to lower levels. This example shows how the various processes compete between themselves in the establishment of statistical equilibrium. Our further calculations of the line profiles showed that the abundance corrections for this star are negative and small.

Despite the multitude of papers devoted to non-LTE studies, our results can be compared only with those from Lind et al. (2011). This is because the collisions with hydrogen atoms were taken into account only in the above paper based on the paper by Barklem et al. (2010). We compared the non-LTE corrections for the MARCS model with the atmospheric
parameters $T_{\rm{eff}}$=5000K, log$\, \textsl{g}$=2.0, [Fe/H]=$-$2 and $\xi_t$=2.0 km/s (Table \ref{tab1}). The agreement is within 0.02.

\section{Analysis of the Solar Na~I Lines}

We used the Sun as a standard star to analyze the abundances of all the subsequent stars. The solar
Na~I lines were analyzed using the Solar Flux Atlas (Kurucz et al. 1984). The set of sodium lines with information about the transitions, the adopted oscillator strengths (log ${gf}$), and the constants for the Stark (log C$_4$) and van der Waals (log C$_6$) interactions is given in Table \ref{tab2}. To calculate the profiles of the line broadening caused by elastic collisions with neutral hydrogen (van der Waals broadening), we used data from Anstee and O$'$Mara (1995) and Barklem and O$'$Mara (1997). For the three weak sodium 6154, 6161, and 5149 \,\AA\, lines, there are no accurate data for
calculating the constant C$_6$. Therefore, we took the same values for them as those calculated for the 5682 and 5688 \,\AA\, lines.

 \begin{table}[htbp]
  \begin{center}
  \caption{ Comparison of our non-LTE corrections with those from Lind et al. (2011) for the selected Na~I lines in
the model atmosphere of a low-metallicity giant with $T_{\rm{eff}}$=5000K, log$\, \textsl{g}$=2.0, [Fe/H]=$-$2 and $\xi_t$=2.0 km/s}
  \label{tab1} 
 {
  \begin{tabular}{c|c|c|c}\hline 
$\lambda$, \,\AA\  & Transition & Lind et al. (2011) & This study \\ \hline
5895.92 &  3s-3p & -0.50 & -0.51 \\
5688.20 &  3p-4d & -0.10 & -0.10 \\
6160.75 &  3p-5s & -0.08 & -0.10 \\
8183.26 &  3p-3d & -0.30 & -0.32\\
 \hline
  \end{tabular}
  }
 \end{center}
\vspace{1mm}
\end{table}

The calculations were performed for two different solar model atmospheres with $T_{\rm{eff}}$=5777 K, 
log$\, \textsl{g}$=4.44, [Fe/H]=0 и $\xi$=0.9 km/s. The first model was computed as described in the $"$Atmospheric Parameters$"$ Section; this is the so-called ATLAS9 model that we will call below the standard model. The second model is MARCS (Gustafsson et al. 2008).

The sodium abundance was determined by analyzing the line profiles. As a rule, the uncertainty in the procedure of describing the theoretical profile is less than 0.02 dex for weak lines and 0.03 dex for strong ones. We convolved the theoretical profiles to take into account the broadenings by rotation with a
velocity of 1.8 km/s and macroturbulence.

 \begin{table}[h]
  \begin{center}
  \caption{Atomic data used for the lines and the solar non-LTE abundance derived by using Kurucz$'$s (column (a))
and MARCS (column (b)) models. The results of test calculations using $\Delta$log C$_6$=-0.3 (c), the rate coefficients for
collisions from Park (1971) (d) and Igenbergs et al. (2008) (e) are presented. The means are given without allowance for
the resonance lines}\label{tab2}{
  \begin{tabular}{c|c|c|c|c|c|c|c|c|c}\hline 
%\multirow{2}{*}{$\lambda$, \,\AA\} & \multirow{2}{*}{Transition} & \multirow{2}{*}{log $gf$} &\multirow{2}{*}{log C$_4$} &
%\multirow{2}{*}{log C$_6$} & \multicolumn{5}{c}{log$\epsilon$ $_{\odot Na}$} \\
$\lambda$, \,\AA\ & Transition & log $gf$ & log C$_4$ & log C$_6$ & \multicolumn{5}{c}{log$\epsilon$ $_{\odot Na}$} \\
    &                      &                      &
    &                      & (a)& (b) & (c) & (d) & (e)\\\hline 
5889.95 &  3s$^{2}$S -- 3p $^{2}$P$^{o}$  & 0.109    & -15.11  &  -31.131 & 6.17 & 6.17 & 6.23 & 6.16  & 6.17 \\
5895.92 &  3s$^{2}$S -- 3p $^{2}$P$^{o}$  & -0.194   & -15.11  &  -31.131 & 6.17 & 6.17 & 6.25 & 6.16  & 6.17 \\
8183.26 &  3p $^{2}$P$^{o}$ -- 3d$^{2}$D  & 0.237    & -13.76  &  -30.407 & 6.18 & 6.19 & 6.23 & 6.18  & 6.18\\
8194.80 &  3p $^{2}$P$^{o}$ -- 3d$^{2}$D  & 0.538    & -13.76  &  -30.407 & 6.17 & 6.17 & 6.22 & 6.17  & 6.17 \\
5682.64 &  3p $^{2}$P$^{o}$ -- 4d$^{2}$D  & -0.706   & -12.18  &  -29.469 & 6.16 & 6.16 & 6.19 & 6.15  & 6.16 \\
5688.20 &  3p $^{2}$P$^{o}$ -- 4d$^{2}$D  & -0.406   & -12.18  &  -29.469 & 6.15 & 6.14 & 6.22 & 6.15  & 6.15 \\
6154.23 &  3p $^{2}$P$^{o}$ -- 5s$^{2}$S  & -1.547   & -14.63  &  -29.469 & 6.23 & 6.23 & 6.23 & 6.22  & 6.23 \\
6160.75 &  3p $^{2}$P$^{o}$ -- 5s$^{2}$S  & -1.246   & -14.63  &  -29.469 & 6.23 & 6.23 & 6.23 & 6.22  & 6.23 \\
5148.84 &  3p $^{2}$P$^{o}$ -- 6s$^{2}$S  & -2.090   & -12.55  &  -29.469 & 6.25 & 6.25 & 6.25 & 6.24  & 6.25 \\ \hline
Average &         &          &         &          & 6.20 & 6.20  & 6.22 & 6.19 & 6.20 \\
Error &         &          &         &          & 0.04 & 0.04  & 0.02 & 0.04 & 0.04 \\\hline 
  \end{tabular}
  }
 \end{center}
\vspace{1mm}
\end{table}

The results obtained with the two models are in good agreement (see Table \ref{tab2}). The mean sodium abundance was obtained from all lines, except the resonance ones, in the scale log $\epsilon_{Н}$ = 12: $\epsilon_{Na}$ = $6.20\pm0.04$. This result agrees well with the results from Asplund et al. (2009), who determined log~$\epsilon_{Na}$ = $6.24\pm0.04$.

We checked the influence of the accuracy of atomic data on the abundance determination. For this purpose, we determined the sodium abundances on the Sun with the standard model atmosphere using various data for collisional processes with electrons: the empirical rate coefficients from Park (1971) and the analytical functions from Igenbergs et al. (2008). The
results agree within 0.01 dex (Table \ref{tab2}).

\section{ANALYSIS OF THE SODIUM ABUNDANCE}

\subsection{Observational Data and the Sample of Stars}

The spectroscopic observations of the stars were carried out with various instruments (Table \ref{tab3}). In addition, we used the archives of the European Southern Observatory (Chile, Paranal, the VLT/UVES spectrograph, program ID 266.D-5655(A)). All spectra were taken with a high spectral resolution (R = $\lambda$/$\delta\lambda$) and a signal-to-noise ratio of at least 100.
All of the data obtained were reduced in the MIDAS standard package, with the exception of those obtained with the VLT (the UVES standard data reduction system) and Otto Struve (IRAF) telescopes.

The stellar sample consists of 79 red giants in the solar neighborhood. The ages, masses, and kinematic parameters, namely the space velocity components (U , V , W ), were taken from Antipova et al. (2003, 2005), Pakhomov et al. (2009a, 2009b, 2011), and Pakhomov (2012, 2013). All stars were separated into three kinematic groups: the thin disk (among which there are 15 Ba~II stars), the thick disk, and the Hercules-stream
stars. When the stars were separated into groups, the following factors were taken into account: the Galactic velocity components, the orbital eccentricity, and the maximum distance of a star from the Galactic
plane. It should be kept in mind that the thin and thick disks have some region of overlapping in kinematics
and metallicity. Therefore, the membership of stars in the thick or thin disk was determined with some
probability by the technique described in Mishenina et al. (2004).

 \begin{table}[h]
  \begin{center}
  \caption{Information on the observed stars}
  \label{tab3} 
 {
  \begin{tabular}{llllccc}\hline 
{$\No$} & {Observatory, country} & {Telescope} & {Spectrograph} & { R } & { $\Delta\lambda$, \,\AA\,} &{Years} \\ \hline
1 & CrAO, Ukraine & 2.6-m MTS & ASP-14 & 40000 & 5100-6800 & 1998-2005\\
2 & Xinglong, China & 2.6-m & Red brunch & 50000 & 5600-9300 & 2006-2011\\
3 & Terskol, Russia & 2-m RCC Karl Zeiss & MAESTRO & 43000 & 3500-9800 & 2006-2010 \\
4 & SAO, Russia & 6-m BTA & NES & 60000 & 5000-6800 & 2000-2003 \\
5 & ESO, Chile & 8-m VLT & UVES & 80000 & 3030-10400 & archive \\
6 & McDonald, USA & 2.1-m Otto Struve & CE & 60000 & 5000-6000 & 2005\\ \hline
  \end{tabular}
  }
 \end{center}
\vspace{1mm}
\end{table}

The stellar age can serve as an additional criterion for the membership of stars in the thick or thin disk, given that the thick-disk stars cannot be younger than 8 Gyr (Fuhrmann 1998). Whereas this can be used for dwarf stars, the age cannot be determined with a satisfactory accuracy for giant stars located very closely on the Hertzsprung--Russell diagram.
Therefore, here we introduced a different criterion, the ratio of $\alpha$-process elements ($\alpha$-elements) to iron. As
numerous studies show, [$\alpha$/Fe] is, on average, higher
for thick-disk stars ([$\alpha$/Fe] > 0.2) than for thin-disk
ones (see, e.g., Bensby et al. 2003), where [$\alpha$/Fe] is
the mean of [Mg/Fe], [Ca/Fe], and [Si/Fe] or [$\langle$Mg+Ca+Si$\rangle$/Fe]. This criterion was applied if the membership probability of a star in the thick disk according to the kinematic criterion was less than 0.99 but no
less than 0.8. As a result, an additional group of 13 stars, the so-called transitional ones, that cannot
be attributed neither to the thick disk nor to the thin one was formed.

\begin{figure}[h]
\setcaptionmargin{5mm}
\onelinecaptionsfalse
\includegraphics[scale=0.325]{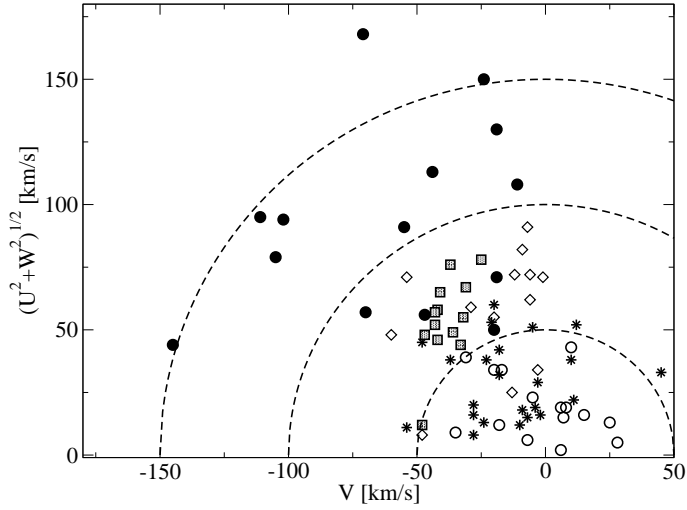}
\captionstyle{normal}
\caption{Toomre diagram for the investigated stars. The thin-disk stars are highlighted by the asterisks, the thin-disk Ba~II stars -- white circles, the thick-disk stars -- black circles, the transitional stars -- diamonds, and the Hercules-stream stars -- gray squares.
}
\label{Kinemat}
\end{figure}

\begin{figure}[h]
\setcaptionmargin{5mm}
\onelinecaptionsfalse
\includegraphics[scale=0.325]{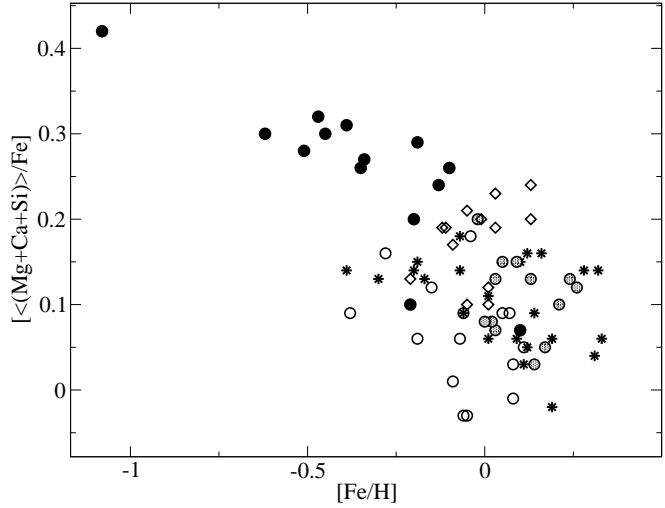}
\captionstyle{normal}
\caption{Abundances of $\alpha$-process elements for the sample being investigated: [$\langle$Mg+Ca+Si$\rangle$/Fe] versus [Fe/H]. The designations are the same as those in Fig. \ref{Kinemat}.}
\label{alfa}
\end{figure}

Four groups of stars are shown on the (U$^{2}$+W$^{2}$)$^{1/2}$ versus V diagram (Fig. \ref{Kinemat}): 38 thin-disk stars
(among which there are 15 Ba~II ones), 15 thick-disk stars, 13 Hercules-stream stars, and 13 transitional stars. Figure \ref{alfa} shows the abundances of $\alpha$-elements in the atmospheres of these groups of stars.
Two stars, HD~74387 and HD~24758, that, judging by their kinematics, belong with confidence (with the probability p = 0.99) to the thick disk but exhibit no overabundances of $\alpha$-elements engage our attention.
These may be the so-called runaway stars that were ejected from the thin disk after binary disruption due to a supernova explosion or the dynamical interactions between stars in dense clusters (Zwicky 1957).

\subsection{Atmospheric Parameters}

\begin{figure}[h]
\setcaptionmargin{10mm}
\onelinecaptionsfalse
\includegraphics[scale=0.325]{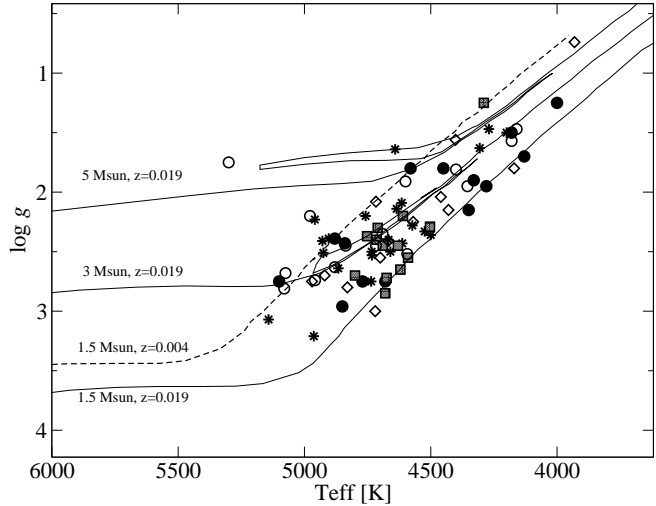}
\captionstyle{normal}
\caption{ Hertzsprung$-$Russell diagram based on the data from Table \ref{tab4} with the evolutionary tracks (Girardi et al. 2000) for initial masses of 1.5, 3.0, and 5.0 $M_\odot$ and solar metallicity (z=0.019) (black solid lines) and the evolutionary track for 1.5 $M_\odot$ with a reduced metallicity (z=0.004) (dashed line).} \label{Param}
\end{figure}

All atmospheric parameters of the stars being investigated, the abundances of $\alpha$-elements ([Mg/H],
[Ca/H], [Si/H]), and the Ba abundance relative to Fe were taken from Antipova et al. (2003, 2005), Pakhomov et al. (2009a, 2009b, 2011), and Pakhomov (2012, 2013). 
Stellar parameters are in the following ranges: effective temperature 3930 < $T_{\rm{eff}}$ < 5300 K, 
surface gravitiy 0.74 < log$\, \textsl{g}$ < 3.21 and
metallicity -1.08 <[Fe/H]< +0.34 (Рис. \ref{Param}). 

These were determined by the methods based on the analysis of lines for iron-peak elements. The metallicities [Fe/H] were determined by averaging [Fe I/H] and [Fe II/H], where [Fe I/H] and [Fe II/H] are the Fе abundances derived from Fe I
and Fe II lines, respectively. 
The model atmospheres were computed by applying the ATLAS9 code (Kurucz 1993) based on the following assumptions: plane-parallel geometry, chemical homogeneity, hydrostatic equilibrium, LTE, and convective equilibrium without overshooting. The blanketing effect was taken into account via the opacity distribution functions (ODFs, Kurucz 1993).

\subsection{Abundance Determination}

\begin{figure}[h]
\begin{center}
\parbox{0.8\linewidth}{\includegraphics[scale=0.325]{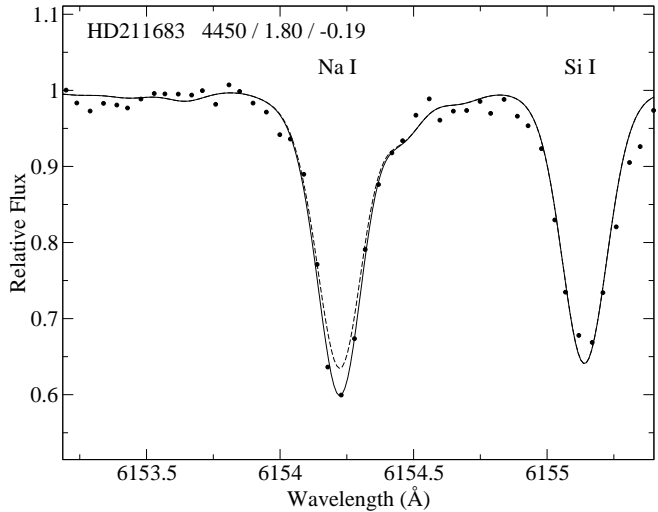}\\
\centering}
\hspace{1\linewidth}
%\hfill
%\\[7ex]
%\parbox{0.8\linewidth}{\includegraphics[scale=0.3]{HD80966.eps}\\
%\centering}
%\hspace{1\linewidth}
%\hfill
%\\[0ex]
\end{center}
\caption{The observed Na~I 6154 \,\AA\, line profile for HD~211683 (black points) in comparison with the non-LTE (solid curve) and LTE (dashed curve) profiles. The LTE profile is drawn at the same sodium abundance as the non-LTE profile. }
\label{Synth}
\end{figure}

The sodium abundance for each star was derived by fitting the synthetic spectra with the observed profiles of two lines, Na~I 6154 and 6161 \,\AA\,. Figure \ref{Synth} shows the observed Na~I 6154 \,\AA\, line profile for HD 211683 that is compared with the non-LTE and LTE line profiles. The deduced abundances are given in Tables \ref{tab4}. The non-LTE [Na/Fe] abundances turn out to be lower than the LTE abundances, on average, by 0.06 dex. Applying the non-LTE approach leads to a decrease in the error of [Na/Fe] for all groups, arguing for it. This is most obvious for the group of transitional stars.

\begin{figure}[h]
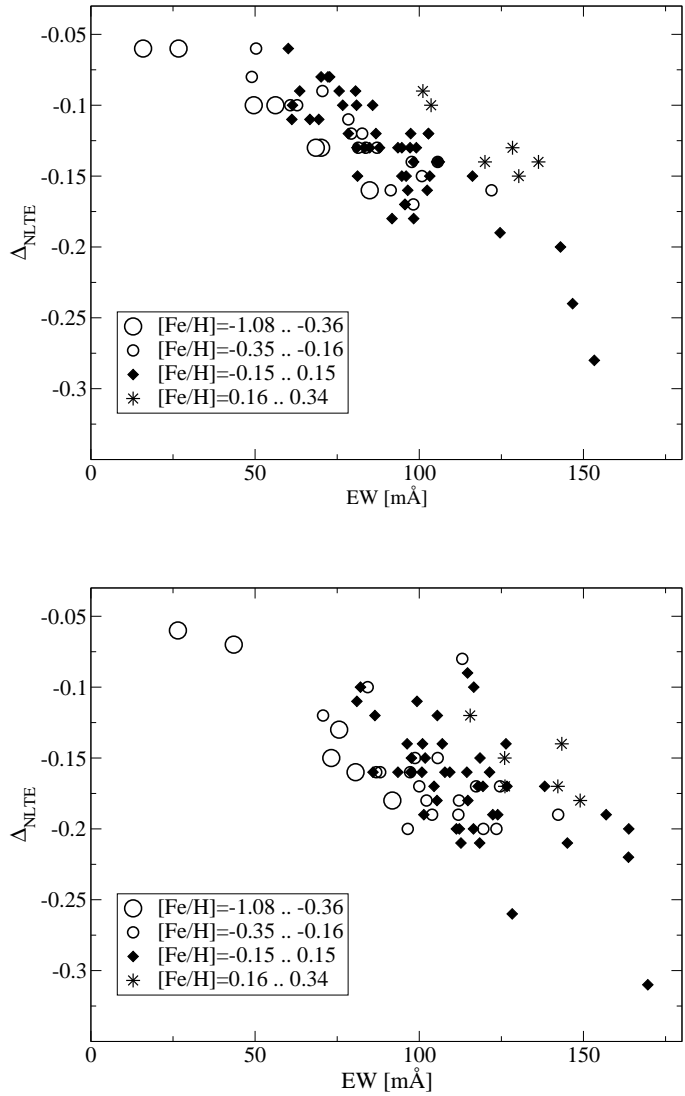

\begin{center}
\parbox{0.8\linewidth}{\includegraphics[scale=0.325]{FIGURE7.eps}\\
\centering}
\hspace{1\linewidth}
\hfill
\\[7ex]
\parbox{0.8\linewidth}{\includegraphics[scale=0.325]{FIGURE8.eps}\\
\centering}
\hspace{1\linewidth}
\hfill
\\[0ex]
\end{center}
\caption{Non-LTE abundance corrections for Na~I 6154 \,\AA\, (top panel) and Na~I 6161 \,\AA\, (bottom panel) in the investigated stellar sample.}
\label{Corrs}
\end{figure}

Figure \ref{Corrs} shows the dependence of the non-LTE corrections on metallicity and line equivalent width (EW). For our sample of stars, the non-LTE abundance corrections for two lines, 6154 and 6161 A, are from $-$0.06 to $-$0.24 dex, depending on the stellar parameters. These are small if EW < 50 m\,\AA\, and become larger with increasing equivalent width. The dependence of the non-LTE corrections on equivalent width is attributable to different line positions on the curve of growth. A comparative analysis of our stars shows that a decrease in metallicity leads to an increase in non-LTE corrections. This is because the electron number density decreases with decreasing metallicity and, hence, the collisional rates are reduced compared to the radiative ones. As a result, the departures from LTE for two stars with identical
equivalent widths will be stronger where the metallicity is lower. On the other hand, the Na~I lines are enhanced with increasing metallicity in such a way that they their formation moves to higher atmospheric layers, where the departures from LTE are stronger. However, the first mechanism turns out to be dominant in the atmospheres of our stars.

\begin{longtable}{ccccccccccccc}
\caption{Adopted atmospheric and kinematic parameters of the stars: the HD number, effective temperature, surface
gravity, iron abundance, microturbulence, mass, Galactic velocity vector components U , V , W relative to the local
standard of rest, membership probability in a given type of population, $\alpha$-element abundances, LTE and non-LTE
[Na/Fe] abundances. Each group of stars, except for the Hercules-stream stars, was divided into two subgroups with
log$g$ < 2.4 and log$g$ > 2.4. The mean [Na/Fe] and its rms deviation are given at the end of each subgroup. For the
Hercules-stream stars, the mean [Na/Fe] is given without allowance for the star HD 10550}\label{tab4} \\
\noalign{\smallskip} \hline \noalign{\smallskip} 
HD &  $T_{\rm{eff}}$ & log$g$ & [Fe/H]  & $\xi$   &   $M$       & U      & V      & W     &  p  & [$\alpha$/Fe] & [Na/Fe] & [Na/Fe] \\
     &  [K]      &[cgs] & [km/s]       &   &   [$M_\odot$] &  [Gyr] & [km/s] & [km/s] & [km/s] &  LTE  &  non-LTE \\\hline
     \endfirsthead 
     
\multicolumn{13}{c}%
{{\bfseries \tablename\ \thetable{} -- continued from previous page}} \\
\hline HD &  $T_{\rm{eff}}$ & log$g$ & [Fe/H]  & $\xi$   &   $M$       & U      & V      & W     &  p  & [$\alpha$/Fe] & [Na/Fe] & [Na/Fe] \\
&  [K]      &[cgs] & [km/s]       &   &   [$M_\odot$] &  [Gyr] & [km/s] & [km/s] & [km/s] &  LTE  &  non-LTE \\ \hline 
\endhead
\hline
 \multicolumn{13}{c}{{Continued on next page}} \\ 
\endfoot     
     
%\noalign{\smallskip} \hline \noalign{\smallskip}
%\endhead 

%\hline
%\endfoot

\hline\hline
\endlastfoot         
         \multicolumn{13}{c}{THIN DISK STARS}      \\
161587  & 4270  &  1.47 &  0.12  &  1.7  &   $3.5\pm0.5$   &   -12   &  -10      &    2   &  0.99       & 0.05   &  0.30  &   0.22    \\
320868  & 4200  &  1.50 & -0.20  &  1.3  &   $2.2\pm2.0$   &    0    &   45      &  -33   &  0.91       & 0.14   &  0.14  &   0.07    \\
 72604  & 4310  &  1.63 & -0.19  &  1.4  &   $1.1\pm0.3$   &   -6    &  -7       &  -14   &  0.99       & 0.15   &  0.13  &   0.06    \\
184938  & 4640  &  1.64 & -0.39  &  1.4  &   $1.8\pm0.3$   &   -38   &   10      &   1    &  0.99       & 0.14   &  0.31  &   0.23    \\
171767  & 4620  &  2.09 &  0.28  &  1.5  &   $2.8\pm0.4$   &   -8    &  -24      &  -10   &  0.99       & 0.14   &  0.39  &   0.32    \\
108123  &  4640 &  2.14 &  0.14  &  1.3  &   $2.9\pm0.3$   &    51   &    -5     &   -1   &  0.99       & 0.09   &  -0.02 &  -0.08    \\
105771  & 4760  &  2.20 & -0.17  &  1.3  &   $2.3\pm0.4$   &   -37   & -23       &  -8    &  0.99       & 0.13   & -0.03  &  -0.04    \\
162391  & 4960  &  2.23 &  0.19  &  1.7  &   $4.1\pm0.5$   &   -15   &  -2       &   -5   &  0.99       & 0.06   &  0.36  &   0.18    \\
179691  &  4570 &  2.28 &  0.01  &  1.3  &   $1.0\pm0.3$   &    52   &  -20      &   -29  &  0.90       & 0.11   &  0.17  &   0.11    \\
107325  & 4520  &  2.33 &  0.12  &  1.2  &   $1.5\pm0.1$   &    1    &  -28      &  -16   &  0.95       & 0.16   &  0.17  &   0.08    \\
 20893  & 4500  &  2.36 &  0.33  &  1.4  &   $2.8\pm0.3$   &    11   &   11      &   -19  &  0.95       & 0.06   &  0.18  &   0.13    \\
162587  & 4900  &  2.39 &  0.09  &  1.6  &   $4.7\pm1.0$   &   -18   &  -4       &   -5   &  0.99       & 0.06   &  0.22  &   0.16    \\
172190  & 4670  &  2.40 &  0.10  &  1.4  &   $2.2\pm0.3$   &    17   &  -28      &  -11   &  0.99       & 0.15   &  0.10  &   0.05    \\
         \multicolumn{11}{l}{ Average } & 0.19 & 0.11     \\
                \multicolumn{11}{l}{ Error } & 0.13 & 0.11     \\
104979  &  4930 &  2.41 & -0.30  &  1.1  &   $2.1\pm0.5$   &   -44   &   12      &  -27   &  0.95       & 0.13   &  0.09  &   0.05    \\ 
108381  & 4610  &  2.43 &  0.32  &  1.4  &   $2.6\pm0.1$   &   -8    &  -28      &    2   &  0.95       & 0.14   &  0.24  &   0.17    \\ 
 94600  & 4660  &  2.50 & -0.06  &  1.2  &   $1.7\pm0.3$   &    20   &  -18      &   -37  &  0.91       & 0.09   &  0.12  &   0.05    \\ 
 26162  & 4730  &  2.50 &  0.16  &  1.5  &   $2.5\pm0.2$   &   -36   &  -37      &   13   &  0.95       & 0.16   &  0.11  &   0.05    \\ 
106714  & 4930  &  2.51 & -0.07  &  1.4  &   $2.6\pm0.1$   &   -5    &  -3       &  -29   &  0.95       & 0.14   &  0.12  &   0.07    \\ 
180112  &  4730 &  2.53 &  0.11  &  1.3  &   $1.6\pm0.3$   &    46   &  -21      &  -26   &  0.95       & 0.03   &  0.17  &   0.09    \\ 
 74212  & 4870  &  2.64 &  0.01  &  1.3  &   $2.3\pm0.4$   &   -11   &  -54      &   2    &  0.95       & 0.06   &  0.34  &   0.23    \\ 
109996  & 4740  &  2.75 &  0.31  &  1.2  &   $2.3\pm0.1$   &   -14   &  -18      & -29    &  0.95       & 0.04   & -0.01  &  -0.02    \\ 
185955  & 5140  &  3.07 &  0.19  &  1.3  &   $2.7\pm0.1$   &   -18   &  -9       &  -2    &  0.95       &-0.02   &  0.07  &   0.03    \\ 
191026  &  4960 &  3.21 & -0.07  &  1.2  &   $1.3\pm0.1$   &    44   &  -48      &  -7    &  0.95       & 0.18   &  0.07  &   0.03    \\ 
         \multicolumn{11}{l}{ Average } & 0.13  & 0.08     \\ 
         \multicolumn{11}{l}{Error}    & 0.10  & 0.07     \\  \hline
        \multicolumn{13}{c}{ THIN DISK BA II STARS}      \\     
 20644  & 4160 &   1.47 &  0.08 &   1.3  &    $5.3\pm0.9$    &     18   &   8     &  -7    &  0.99      & -0.01  &  0.52  &  0.41    \\ 
158899  & 4180 &   1.57 & -0.09 &   1.3  &    $2.4\pm0.5$    &    -11   &  7      &  -10   &  0.96      & 0.01   &  0.33  &  0.24    \\ 
204075  & 5300 &   1.75 & -0.06 &   2.2  &    $4.8\pm0.5$    &      9   &   25    &   9    &  0.99      & -0.03  &  0.03  &  0.04    \\ 
  9856  & 4400 &   1.81 & -0.19 &   1.6  &    $2.5\pm0.5$    &    -13   &   15    &  -10   &  0.96      & 0.06   &  0.25  &  0.18    \\ 
199939  & 4600 &   1.91 & -0.38 &   1.7  &    $2.5\pm0.5$    &     33   &  -20    &  -8    &  0.99      & 0.09   &  0.23  &  0.21    \\ 
 83618  & 4360 &   1.95 & -0.04 &   1.5  &    $2.5\pm0.5$    &    26    &  -17    &   22   &  0.95      & 0.18   &  0.09  &  0.00    \\ 
 77247  & 4980 &   2.20 &  0.08 &   1.5  &    $3.9\pm0.5$    &     23   &  -5     &   0    &  0.99      & 0.03   &  0.07  &  0.01    \\ 
176411  & 4690 &   2.35 & -0.06 &   1.4  &    $2.2\pm0.5$    &    -8    &  -35    &   5    &  0.99      & 0.09   &  0.31  &  0.23    \\ 
         \multicolumn{11}{l}{ Average} & 0.23 & 0.17     \\
         \multicolumn{11}{l}{ Error } & 0.16 & 0.14     \\
         \hline \\
181053  & 4840 &   2.45 & -0.15 &   1.3  &    $2.0\pm0.5$    &    -19   &   6     &   2    &  0.99      & 0.12   &  0.10  &  0.10    \\ 
 49293  & 4720 &   2.45 &  0.07 &   1.4  &    $3.3\pm0.5$    &     2    &   6     &  -1    &  0.99      & 0.09   &  0.22  &  0.16    \\ 
153210  & 4590 &   2.52 &  0.11 &   1.1  &    $1.6\pm0.5$    &    -38   &  -31    &   7    &  0.99      & 0.05   &  0.09  &  0.02    \\ 
205011  & 4880 &   2.63 & -0.05 &   1.5  &    $2.4\pm0.5$    &      5   &   28    &   1    &  0.99      & -0.03  &  0.15  &  0.07    \\ 
133208  & 5080 &   2.68 &  0.05 &   1.5  &    $3.5\pm0.5$    &    6     &  -7     &  -1    &  0.99      & 0.09   &  0.25  &  0.19    \\ 
 65854  & 4960 &   2.74 & -0.28 &   1.3  &    $1.6\pm0.5$    &    -10   &  -18    &  -6    &  0.99      & 0.16   &  0.20  &  0.13    \\ 
199394  & 5080 &   2.81 & -0.07 &   1.5  &    $2.3\pm0.5$    &     43   &   10    &  -4    &  0.99      & 0.06   &  0.25  &  0.18    \\ 
         \multicolumn{11}{l}{ Average} & 0.18 & 0.12     \\
         \multicolumn{11}{l}{ Error } & 0.07 & 0.06  \\ \hline       
        \multicolumn{13}{c}{  THICK DISK STARS}      \\    
 37171 &  4000 &   1.25 & -0.62 &   1.4  &    $1.1\pm0.2$   &    -113  &   -19   &    65  &  0.99       & 0.30   &  0.14  &  0.07     \\
141472 &  4180 &   1.50 & -0.39 &   1.4  &    $1.1\pm0.3$   &    155   &   -71   &    64  &  0.99       & 0.31   &  0.04  &  0.00     \\
 10057 &  4130 &   1.70 & -0.35 &   1.4  &    $1.7\pm0.3$   &      4   &   -47   &    56  &  0.97       & 0.26   &  0.06  & -0.01     \\
127243 &  5100 &   2.75 & -0.51 &   1.3  &    $1.6\pm0.3$   &     79   &   -111  &    53  &  0.99       & 0.28   & -0.05  & -0.02     \\
 80966 &  4580 &   1.80 & -1.08 &   1.5  &    $1.2\pm0.3$   &    -23   &   -105  &    76  &  0.99       & 0.42   &  -0.1  & -0.08     \\
180682 &  4330 &   1.90 & -0.45 &   1.4  &    $1.1\pm0.2$   &    -90   &   -11   &    60  &  0.99       & 0.30   &  0.15  &  0.11     \\
  2901 &  4350 &   2.15 & -0.47 &   1.3  &    $0.7\pm0.3$   &     24   &   -145  &   -37  &  0.99       & 0.32   &  0.10  &  0.05     \\
141353 &  4280 &  1.95  & -0.10 &   1.4  &    $2.0\pm0.4$   &   -9     &   -70   &   -56  &  0.97       & 0.26   &  0.08  & -0.02     \\
211683 &  4450 &  1.80  & -0.19 &   1.4  &    $1.9\pm0.3$   &   -7     &   -20   &    50  &  0.80       & 0.29   &  0.16  &  0.11     \\
           \multicolumn{11}{l}{ Average } & 0.06   & 0.02     \\
              \multicolumn{11}{l}{ Error } & 0.09 & 0.07     \\
 74387 &  4840 &   2.43 & -0.21 &   1.3  &    $2.3\pm1.2$   &    -94   &   -44   &   -63  &  0.99       & 0.10   &  0.11  &  0.07     \\
 24758 &  4680 &   2.75 &  0.10 &   1.2  &    $1.4\pm0.2$   &    -124  &   -24   &   -84  &  0.99       & 0.07   &  0.17  &  0.12     \\
203344 &  4770 &   2.75 & -0.13 &   1.3  &    $1.8\pm0.2$   &     62   &   -102  &    -70 &  0.99       & 0.24   &  0.08  &  0.08     \\
   249 &  4850 &   2.96 & -0.20 &   1.2  &    $1.5\pm0.2$   &     45   &   -55   &   -79  &  0.99       & 0.20   &  0.08  &  0.08     \\
106398 & 4880  &  2.39  & -0.34 &   1.3  &    $2.5\pm0.5$   &    28    &   -19   &    65  &  0.90       & 0.27   &  0.04  &  0.04     \\
           \multicolumn{11}{l}{Average } & 0.10 & 0.08    \\   
            \multicolumn{11}{l}{ Error} & 0.05 & 0.03     \\ \hline    
        \multicolumn{13}{c}{ HERCULES STREAM}      \\  
 10550  & 4290  &  1.25 &  0.03  &  1.5  &    $7\pm1$     &   -58     &   -42    &    3   & 0.58       &0.07    &  0.66  &   0.52   \\ 
203504  & 4610  &  2.20 &  0.03  &  1.4  &   $1.9\pm0.2$   &    -40    &   -47    &   27   & 0.82       &0.13    &  0.02  &  -0.03   \\
 49520  & 4500  &  2.29 &  0.26  &  1.5  &   $2.8\pm0.3$   &    -64    &   -41    &  -11   & 0.88       &0.12    &  0.11  &   0.03   \\
188853  & 4710  &  2.30 &  0.02  &  1.4  &   $1.9\pm0.3$   &    -10    &   -48    &   -7   & 0.85       &0.08    &  -0.06 &  -0.08   \\
 12139  & 4750  &  2.37 & -0.02  &  1.5  &   $2.2\pm0.2$   &    -46    &   -43    &   24   & 0.83       &0.20    &  0.05  &   0.02   \\
 10437  & 4710  &  2.40 &  0.17  &  1.3  &   $2.3\pm0.3$   &    -41    &   -33    &  -16   & 0.81       &0.05    &  -0.05 &  -0.10   \\
115136  & 4630  &  2.45 &  0.13  &  1.3  &   $1.8\pm0.2$   &    -75    &   -37    &   15   & 0.85       &0.13    &  0.21  &   0.14   \\
221742  & 4690  &  2.45 &  0.05  &  1.4  &   $1.9\pm0.2$   &    -44    &   -42    &   15   & 0.87       &0.15    &  -0.04 &  -0.06   \\
115061  & 4590  &  2.55 &  0.09  &  1.3  &   $1.3\pm0.2$   &    -66    &   -31    &   14   & 0.84       &0.15    &  0.15  &   0.04   \\
 94669  & 4620  &  2.65 &  0.00  &  1.3  &   $1.5\pm0.4$   &    -42    &   -43    &  -38   & 0.41       &0.08    &  -0.01 &  -0.09   \\
180314  & 4800  &  2.70 &  0.21  &  1.3  &   $2.4\pm0.2$   &    -42    &   -36    &  -26   & 0.86       &0.10    &  0.24  &   0.13   \\
 55280  & 4680  &  2.72 &  0.24  &  1.3  &   $1.9\pm0.2$   &    -52    &   -32    &  -17   & 0.85       &0.13    &  0.02  &  -0.06   \\
184423  & 4680  &  2.85 &  0.14  &  1.3  &   $1.7\pm0.2$   &    -78    &   -25    &   -7   & 0.80       &0.03    &  0.09  &   0.02   \\
           \multicolumn{11}{l}{Average } & 0.06 & 0.00    \\ 
           \multicolumn{11}{l}{ Error } & 0.10 & 0.08     \\ \hline 
        \multicolumn{13}{c}{ transition objects }      \\ 
183400  & 3930  &  0.74 &  0.13  &  1.5  &   $1.9\pm1.1$  &    24     &  -13     &   8    &             & 0.24   &  0.42  &   0.22    \\
 109519 & 4400  &  1.56 &  0.03  &  1.4  &   $2.6\pm0.5$  &    28     &  -3      &  -19   &             & 0.23   &  0.10  &   0.06    \\
152879  & 4170  &  1.80 &  0.01  &  1.4  &   $1.4\pm0.5$  &    10     &   -29    &   58   &  0.04/0.95  & 0.12   &  0.26  &   0.15   \\ 
162113  & 4460  &  2.04 &  0.13  &   1.4 &   $1.3\pm0.2$  &    -71    &   -1     &    5   &             & 0.20   &  0.07  & -0.01    \\ 
105475  & 4720  &  2.08 & -0.09  &  1.3  &   $2.3\pm0.3$  &    71     &   -6     &   10   &             & 0.17   &  0.09  &   0.08    \\
 92095  & 4430  &  2.15 & -0.05  &  1.4  &   $1.9\pm0.3$  &    68     &  -54     &   22   &  0.16/0.83  & 0.10   &  -0.01 &  -0.05   \\ 
197752  & 4570  &  2.25 &  0.01  &  1.3  &   $2.6\pm0.5$  &   -79     &  -7      &  -45   &  0.41/0.58  & 0.10   &  0.18  &   0.08   \\ 
           \multicolumn{11}{l}{ Average } & 0.16 & 0.08    \\ 
           \multicolumn{11}{l}{ Error } & 0.14 & 0.09    \\ 
 74442  & 4670  &  2.41 & -0.01  &  1.5  &   $2.1\pm0.1$  &   -3      &  -48     &  -7    &             & 0.20   &  0.08  &   0.03    \\
212074  & 4700  &  2.55 &  0.03  &  1.3  &   $2.3\pm0.3$  &   -6      &  -6      &   62   &  0.03/0.96  & 0.19   &  0.07  &   0.00   \\ 
100696  & 4920  &  2.70 & -0.21  &  1.3  &   $2.4\pm0.3$  &   -41     &  -20     &   36   &  0.52/0.47  & 0.13   &  0.02  &   0.00   \\ 
 94860  & 4970  &  2.75 & -0.05  &  1.2  &   $2.8\pm0.3$  &    20     &  -60     &  -44   &  0.15/0/80  & 0.21   &  0.31  &   0.25   \\ 
104985  & 4830  &  2.80 & -0.11  &  1.3  &   $1.9\pm0.3$  &   -76     &  -9      &   30   &  0.38/0.61  & 0.19   &  -0.01 &  -0.05   \\ 
  6555  & 4720  &  3.00 & -0.12  &   1.2 &   $1.2\pm0.2$  &   -25     &   -12    &  -68   &  0.15/0.84  & 0.19   &  0.19  &  0.09    \\ 
             \multicolumn{11}{l}{ Average} & 0.11 & 0.05    \\ 
                  \multicolumn{11}{l}{Error } & 0.12 & 0.10     \\ 
\end{longtable}

\begin{table}[h]
  \begin{center}
  \caption{Sensitivity of the [Na/Fe] abundance to the atmospheric parameters and log C$_6$ for 
HD 162587 (EW(6154)=94 m\AA\,, EW(6161)=115 m\AA\,, $T_{\rm{eff}}$=4900 K, log$g$=2.39, $\xi$=1.6 km/s) и
HD 10550 (EW(6154)=147 m\AA\,, EW(6161)=164 m\AA\,, $T_{\rm{eff}}$=4290 K, log$g$=1.25, $\xi$=1.5 km/s)
}
  \label{tab5} 
 {
 %\scriptsize
  \begin{tabular}{llccccc}\hline 
{ Звезда } & {        }  & {$\Delta$$T_{\rm{eff}}$=80 K} & {$\Delta$ log$g$=0.15} & {$\Delta$ $\xi$=0.15 км/с} & {$\Delta$log C$_6$=-0.3} & Total \\ \hline
HD 162587   & $\Delta$ [Na/Fe I]  & 0.00  &  0.00 & 0.02    & 0.00   & 0.02  \\
            & $\Delta$ [Na/Fe II] & 0.11  & -0.08 & 0.03    & 0.00   & 0.14  \\ 
HD 10550    & $\Delta$ [Na/Fe I]  & 0.08  & -0.03 & -0.02   & 0.00   & 0.09  \\
            & $\Delta$ [Na/Fe II] & 0.24  & -0.08 & -0.01   & 0.00   & 0.25 \\ \hline            
  \end{tabular}
  }
 \end{center}
\vspace{1mm}
\end{table}

The Na~I abundance on the Sun was taken from our analysis of the solar Na~I 6154 and 6161 \,\AA\, lines
in this paper (see above) and is log log$\epsilon$~$_{\odot Na}$=6.23 for non-LTE and log$\epsilon$~$_{\odot Na}$=6.33 in LTE. The [Na/Fe] abundance determination depends on the accuracy of the atmospheric parameters. We calculated the uncertainty in the sodium-to-iron abundance ratios derived from neutral ([Na/Fe~I]) and ionized ([Na/Fe~II])
iron lines as a function of the probable errors in the atmospheric parameters and in the broadening
parameter $\Delta$log C$_6$=$-$0.3 for two stars, HD 162587 (EW(6154) = 94 m\,\AA\,, EW(6161) = 115 m\,\AA\,) and 
EW(6161) = 164 m\,\AA\,) and HD 10550 (EW(6154) = 147 m\,\AA\,, (Table \ref{tab5}). The total error was calculated by assuming all of the individual errors to be independent. The errors in [Na/Fe I] due to the uncertainties in $T_{\rm{eff}}$, log$\, \textsl{g}$ are small, because Na~I and Fe I have the same sensitivity to a change in parameters. The Fe II lines are very sensitive to the error in log g, while the Na~I lines are insensitive. On the other hand, the Na~I lines will be affected by a change in $T_{\rm{eff}}$, while the Fe II lines will not be affected. As a result, Na~I and Fe~II behave differently as the parameters change. The estimated errors are typical of all stars with similar equivalent
widths.

\section{DISCUSSION}

\subsection{Analysis of Results}

\begin{figure}[h]
\setcaptionmargin{5mm}
\onelinecaptionsfalse
\includegraphics[scale=0.325]{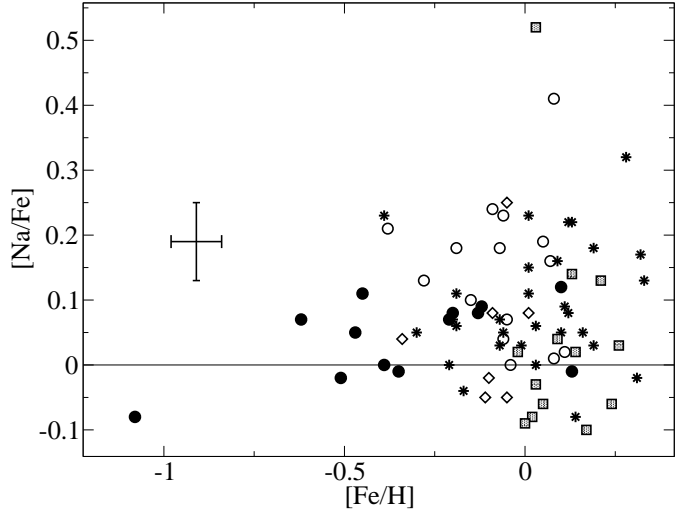}
\captionstyle{normal}
\caption{[Na/Fe] abundance versus [Fe/H]. The designations are the same as those in Fig. \ref{Kinemat}. The error bar is indicated on the left.}
\label{Metal}
\end{figure}

The sodium abundances with respect to iron are shown in Figure \ref{Metal} for our sample of stars. We do
not observe any dependence of [Na/Fe] on metallicity for the kinematic groups. This is most likely attributable to the narrow metallicity range for the stars being investigated. The following abundances were derived:
$\langle$[Na/Fe]$\rangle$ = $\bf0.10\pm0.10$ for the thin-disk stars, [Na/Fe] = $\bf0.04\pm0.06$ for the thick-disk
stars, [Na/Fe] = $\bf0.00\pm0.08$ for the Hercules-stream stars (without HD 10550), $\langle$[Na/Fe]$\rangle$ = $\bf0.15\pm0.11$ for Ba~II stars, and $\langle$[Na/Fe]$\rangle$ = $\bf0.07\pm0.10$ for the transitional stars.
The derived values of [Na/Fe] are equal to their solar values, within the error limits.
This suggests that once the [Fe/H] abundance in the Galaxy had reached a value of about $-$0.5, the sodium
synthesis mechanisms did not undergo any changes.

\begin{figure}[h]
\setcaptionmargin{5mm}
\onelinecaptionsfalse
\includegraphics[scale=0.325]{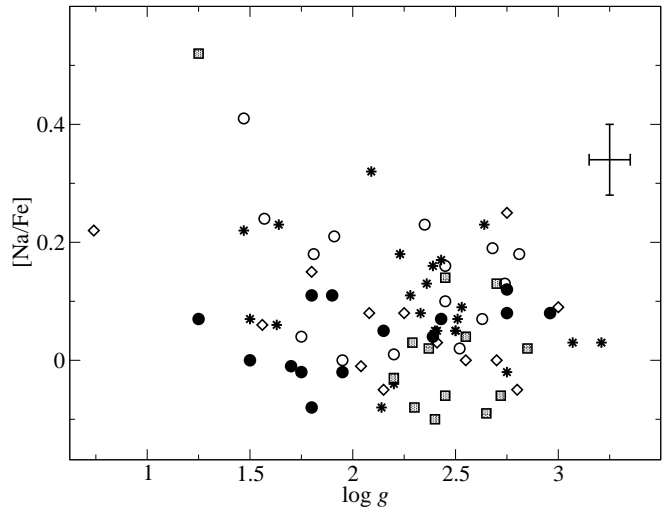}
\captionstyle{normal}
\caption{[Na/Fe] abundance versus log${g}$. The designations are the same as those in Fig. \ref{Kinemat}. The error bar is indicated on the right.}
\label{logg1}
\end{figure}

On the other hand, we are interested in whether
the [Na/Fe] abundance actually increased in our stars
during their evolution owing to the NeNa cycle and
sodium was brought to the surface through large-scale convection. If this is true, then we must see
a correlation between [Na/Fe] and luminosity or an
anticorrelation between [Na/Fe] and log$g$. In Figure \ref{logg1},
[Na/Fe] is plotted against log$g$. For a finer analysis,
we separated the stars into two groups with log$g$<2.4 and log$g$>2.4 following Cayrel et al. (2004). The
following results were obtained:

--- for stars with log$g$>2.4, we found no differences between the thin-disk ($\langle$[Na/Fe]$\rangle$=$\bf0.08\pm0.07$) and thick-disk ($\langle$[Na/Fe]$\rangle$=$\bf0.08\pm0.03$) stars;
 
--- for stars with log$g$<2.4, we detected a slight
increase in Na with respect to Fe, on average, for the
thin-disk stars ($\langle$[Na/Fe]$\rangle$=$\bf0.11\pm0.11$) compared
to the thick disk stars ($\langle$[Na/Fe]$\rangle$=$\bf0.02\pm0.07$) along with a larger scatter.   

Boyarchuk and Lyubimkov (1981) established the
dependence of [Na/Fe] on log$g$ for F giants for the
first time. Subsequently, Boyarchuk et al. (2001) described an anticorrelation between the [Na/Fe] overabundances and log g for cool giants. Having investigated 177 clump giants with log$g$>2.0, Mishenina
et al. (2006) found no correlation of [Na/Fe] with log$g$.

\begin{figure}[h]
\setcaptionmargin{5mm}
\onelinecaptionsfalse
\includegraphics[scale=0.325]{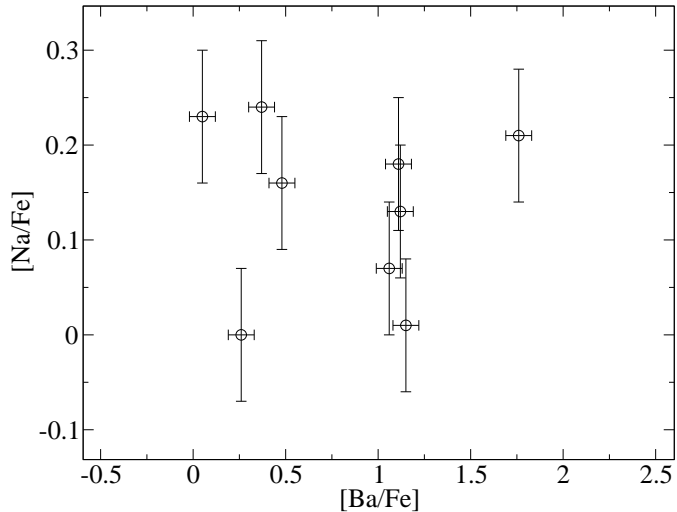}
\captionstyle{normal}
\caption{[Na/Fe] and [Ba/Fe] for nine barium stars.}
\label{Ba}
\end{figure}

Our results reveal no clear correlation between
[Na/Fe] and log g for the sample of thin- and thick-disk stars being investigated. In contrast to the thin-disk stars, the overwhelming majority of thick-disk stars have low masses (M/$M_\odot$ < 2.0). According to
presen-day theoretical views, the NeNa-cycle reactions are inefficient in low-mass (M/$M_\odot$ < 1.5) stars.
As a consequence, no sodium overabundances must be observed.
For 15 thin-disk Ba~II stars, on average, we obtained an enhanced sodium abundance: [$\langle$[Na/Fe]$\rangle$=$\bf0.17\pm0.14$ for stars with log$g$<2.4 and $\langle$[Na/Fe]$\rangle$=$\bf0.12\pm0.06$ for stars with log$g$>2.4. However, it is unlikely that this is related to the Na synthesis in the
$s$-process, because there is no dependence of [Na/Fe]
on [Ba/Fe], as is shown in Figure \ref{Ba} constructed for nine stars from our sample.

The Hercules-stream stars exhibit, on average,
the solar sodium abundance $\langle$[Na/Fe]$\rangle$=$\bf0.00\pm0.08$
if we disregard the star HD~10550, which exhibits
a large overabundance (Table \ref{tab4}) and probably does
not belong to the Hercules stream (p = 0.58). To
all appearances, this star is at a post-main-sequence
evolutionary phase, because it has a high mass and
this explains the large sodium overabundance.

\subsection{Comparison with Other Studies}

Most of the studies on sodium are devoted to
stars in globular and open clusters (Table \ref{tab6}), while
in this paper we investigate field stars. Our LTE
results for the thin-disk stars are in good agreement with the data from Luck and Heiter (2007)
and Takeda et al. (2008) for the thin-disk stars and with the data from Ishigaki et al. (2013) and Alves-
Brito et al. (2010) for the thick-disk stars. However, our LTE [Na/Fe] abundances for the thin-disk stars
are, on average, higher than those from Alves-Brito et al. (2010). Our non-LTE results are consistent
with the [Na/Fe] values for the clump giants from Mishenina et al (2006) and for the thick-disk stars from Ishigaki et al. (2013). In the paper by Mishenina et al. (2006), we found three stars common to our paper. For HD 108381, log$\epsilon_{Na}$ = 6.73 in our paper and log$\epsilon_{Na}$ = 6.71 in their paper, indicative of good
agreement. On the other hand, [Na/Fe] for the same
star is 0.29 in Mishenina et al. (2006) and 0.18 in
our paper. This [Na/Fe] difference is explained by the
fact that the metallicities found in these two papers differ by 0.13 dex. For HD 196714, log$\epsilon_{Na}$ = 6.19 in
Mishenina et al. (2006) and log$\epsilon_{Na}$ = 6.24 in our paper at identical physical parameters of the star being
investigated. For HD 100696, we obtained log$\epsilon_{Na}$ = 6.03, while Mishenina et al. (2006) found log$\epsilon_{Na}$ = 5.98. In this case, the discrepancy by 0.05 dex may be due to the differences in log g and microturbulence.

\begin{longtable}{rccccccc}
\caption{Overview of the published data on Na abundance determinations for late-type stars, where ОC and GC stand
for open and globular clusters, respectively}\label{tab6}\\
\hline
Reference&  & Number  & Stellar    & Method   &  $\langle$[Na/Fe]$\rangle$  & T$_{eff}$   & log$g$  \\
      & &    of stars  & subsystems  &         &                             & [K] &      \\
\hline

     \endfirsthead 
     
\multicolumn{8}{c}%
{{\bfseries \tablename\ \thetable{} -- continued from previous page}} \\\hline
Reference&  & Number  & Stellar    & Method   &  $\langle$[Na/Fe]$\rangle$  & T$_{eff}$   & log$g$  \\
      & &    of stars  & subsystems  &         &                             & [K] &      \\ \hline
\endhead
\hline
 \multicolumn{8}{c}{{Continued on next page}} \\ 
\endfoot 
\hline\hline
\endlastfoot  
%\endfirsthead 
%\hline
%         &      &      &  & & & &          \\
%\hline
%\endhead
%\hline
%\endfoot
%\hline\hline
%\endlastfoot
         \multicolumn{8}{c}{ GIANTS}      \\
Bragaglia et al. &  2001 & 3    & OC (NGC 6819)    &  LTE                  & 0.47 & 4740-4860 & 2.6-2.7  \\ 
Carretta et al.  &  2003 & 70   & GC (NGC 2808)    &  non-LTE                 & 0.37 & 3970-4900 & 0.6-2.4  \\
                &       &      &                  &  (Gratton, 1999)      &      &           &            \\
                &       &      &                  &  LTE                  & 0.22 &  &    \\
Friel et al.     & 2003  & 4    & OC (Collinder261)&  LTE                  & 0.48 & 4000-4490 & 0.7-2.2   \\
Pasquini et al.  & 2004  & 5    & OC (IC 4651) &  LTE                  & 0.19 & 4900-5800 & 2.7-3.5    \\   
Gratton et al.   & 2005  & 7        & GC (NGC 6752)& non-LTE   & 0.29 & 4850-5030 & 2.0-2.5   \\
                &       &      &                  &  (Gratton, 1999)      &      &           &         \\
Carretta et al.  & 2005  & 6   & OC (Collinder261)  & non-LTE  & 0.33 & 3980-4580 & 0.4-2.1   \\
                &       &      &                  &  (Gratton, 1999)      &      &           &          \\
Yong et al.      & 2005  & 21  & GC (NGC 6752)      & LTE                  & 0.27 & 3890-4950 & 0.3-2.4   \\
Mishenina et al.  & 2006  & 177 & clump giants        & non-LTE                 & 0.10 & 4450-5280 & 1.5-3.2  \\
Jacobson et al.  & 2007  & 17  & OC (NGC 7142)      & LTE                  & 0.56  & 4500-5300 & 2.0-3.0  \\
                &       & 16  & OC (NGC 6939)      & LTE                  & 0.40  & 4200-5100 & 1.5-3.0  \\
                &       & 11  & OC (IC 4756)       & LTE                  & 0.57  & 5000-5100 & 2.0-2.5  \\
Luck and Heiter   & 2007  & 298 & thin               & LTE                  & 0.12  & 4800-5100 & 2.6-2.8 \\
Takeda et al.     & 2008  & 322 & thin  & LTE                  & 0.20  & 4490-5930 & 1.4-3.3  \\
Carretta et al.  & 2009  & 202 & 19 GCs             & non-LTE  & 0.34  & 4060-4800 & 0.4-2.3 \\
                &       &      &                  &  (Gratton, 1999)      &      &           &         \\
Smiljanic et al. & 2009  & 31  & 10 OCs             & non-LTE   & solar & 4360-5130 & 1.8-3.2 \\
                &       &      &                  &  (Takeda, 2003)      &      &           &       \\
Liu et al.       & 2009  & 8   & Ba~II    &  LTE  &    0.00   &  4280-5550  &  1.63-3.64 \\ 
Alves-Brito et al. & 2010 & 29 & thin               & LTE                  & 0.11  & 3800-5000 & 1.0-3.2  \\
                  &      & 22 & thick              & LTE                  & 0.07  & 3900-5000 & 1.0-3.7   \\
Mikolaitis et al.  & 2010 & 6  & OC (NGC 6134)      & non-LTE  & $0.04\pm0.06$ & 4940-5050  & 2.5-3.1 \\
               &       &      &                  &  (Gratton, 1999)      &      &           &         \\
Gratton et al.    & 2011 & 36 & GC (NGC 2808)      & non-LTE  & 0.14  & 5430-5690  & 2.5-2.6 \\
               &       &      &                  &  (Gratton, 1999)      &      &           &           \\
Jacobson et al.    & 2011 & 19 & OC (M67)           & LTE                  & $0.03\pm0.06$ & 4200-5100  & 1.8-3.0 \\
Pereira et al.    & 2011 & 12 & Ba~II        & LTE                  & $0.19\pm0.09$ &  4700-5300 & 2.4-3.2 \\                  
Johnson and Pilachowski    & 2012 & 113& GC (M13)          & LTE                   & [-0.6~0.6] & 4000-4950 & 0.8-2.5  \\
Gratton et al.    & 2012  & 110 & GC (47Tucanae)   & non-LTE  & 0.48 & 4920-5260 & 2.1-2.3  \\
               &       &      &                  &  (Gratton, 1999)      &      &           &         \\
Smiljanic          & 2012  & 4   & OC (Hyades)      & non-LTE      & 0.30 & 4830-4970 & 2.6-2.7 \\
               &       &      &                  &  (Lind 2011)     &      &           &         \\
Ishigaki et al.    & 2013  & 97  & thick, halo       & LTE                  & 0.13& 4000-6600 & 1.3-3.7   \\
                  &       &     &                   & non-LTE   & 0.00& 4000-6600 & 1.3-3.7   \\
                                 &       &      &                  &  (Takeda, 2003)      &      &           &   \\
Kacharov et al.   & 2013  & 15  & GC (M75)          & LTE                  & 0.30& 3690-4230 & 0.3-1.3  \\
Mu$\tilde n$oz et al.      & 2013  & 8   & GC (NGC 3201)     & non-LTE & 0.07 & 4180-4480 & 0.5-1.1   \\
                  &       &      &                  & (Mashonkina, 2000)      &      &           &      \\

\hline
         \multicolumn{8}{c}{DWARFS}      \\                  
Edvardsson et al.  & 1993  & 189 &  field stars
            & LTE                  & $0.08\pm0.05$ &5600-7000  & 3.8-4.5   \\
Prochaska et al.   & 2000  & 10  &  thick            & LTE                  & $0.09\pm0.01$ &5220-5700 & 4.2-4.6  \\
Reddy et al.       & 2003  & 181 &  thin             & LTE                  & $0.07\pm0.04$ & 5500-6500 & 3.4-4.6  \\
                  &       &     &  thick            & LTE                  & $0.10\pm0.04$ &  &    \\
Bensby et al.      & 2003  & 45  &  thin              & LTE                  & $0.07\pm0.09$      & 5020-6475  & 3.7-4.5\\
                  &       & 21  &  thick             & LTE                  & $0.05\pm0.07$      & 5320-6040 & 3.5-4.6 \\
Pasquini et al.   & 2004  & 17  &                   & LTE                  & -0.09 & 5960-6390 & 4.3-4.4     \\
Bensby et al.      & 2005  & 15  & thin              & LTE                  & $0.05\pm0.08$& 5810-6870 & 3.9-4.5 \\
                  &       & 17  & thick             & LTE                  & $0.07\pm0.05$& 5000-6410 & 3.1-4.4 \\
Soubiran and Girard    & 2005  & 7   & Hercules stream   & LTE                  & $0.02\pm0.05$ &   & \\
Mishenina et al.   & 2008  & 131 &  field stars
            & non-LTE                 & $0.02\pm0.15$ & 4200-6000 & 4.0-4.8 \\ 
 \hline
\end{longtable}

While analyzing the thick- and thin-disk red giants, Alves-Brito et al. (2010) found no difference in
[Na/Fe] abundance between the thick- and thin-disk stars. For the thick- and thin-disk dwarfs, Reddy
et al. (2003) and Bensby et al. (2003) found no difference in [Na/Fe] abundance between them either.
Bensby et al. (2005) argue that the thin-disk dwarfs seem more enriched with sodium, but, having calculated the mean, we did not see any difference (Table \ref{tab6}). Comparing our results for the thick and thin disks, we found no differences for giants with log$g$ > 2.4, while for giants with log$g$ < 2.4 the [Na/Fe] abundance in the thin-disk stars is, on average, higher than that in the thick-disk stars.

We used the paper by Mishenina et al. (2008) to compare our giants with field dwarfs, because the
non-LTE approach was applied only in their paper. We found no differences between them.
For the Ba~II and Hercules-stream stars, our paper is the first one where the [Na/Fe] abundance
was determined without assuming LTE. The enhanced LTE [Na/Fe] abundance that we derived in
the Ba~II stars is consistent with the LTE results
from Pereira et al. (2011). The LTE abundances
for the Hercules-stream dwarfs from Soubiran and
Girard (2005) are consistent with our non-LTE
results.

\subsection{Comparison of the Measured Na Abundances
with Theoretical Predictions}

\begin{figure}[h]
\setcaptionmargin{5mm}
\onelinecaptionsfalse
\includegraphics[scale=0.325]{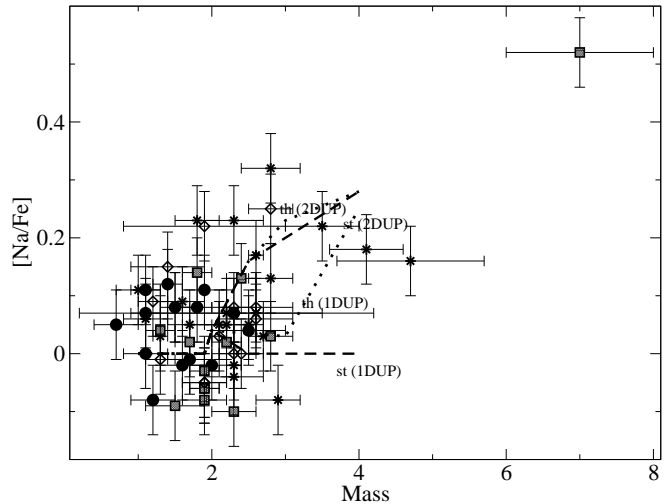}
\captionstyle{normal}
\caption{[Na/Fe] abundance for the stars from the sample being investigated (except for the Ba~II ones) versus stellar mass. The dashed lines indicate the theoretical predictions of the standard theory (st) and the theory with thermohaline instability (th). The designations are the same as those in Fig. \ref{Kinemat}. }
\label{theory}
\end{figure}

There are theories that predict an increase in [Na/Fe] on the stellar surface after the first or second dredge-up (1DUP, 2DUP). During the first dredge-up, the chemical composition on the stellar surface changes when the outer convective zone
propagates deep into the star and reaches the layers that contain the material involved in thermonuclear
reactions (Iben 1967). The change in surface sodium abundance depends on the initial stellar mass and
luminosity. We used the theoretical modeling results that give various predictions of the change in sodium
abundance on the stellar surface until the star begins
to ascend the AGB. The first theory is called the
standard one, in which there is no other mechanism
than convection. The second theory differs from the
standard one in that it takes into account the thermohaline instability (Charbonnel and Lagarde 2010).
The thermohaline instability is the mixing process
that arises as the mean molecular weight of the
stellar matter increases toward the stellar surface.
The [Na/Fe] ratio predicted by the two theories as
a function of the stellar mass is shown in Fig. \ref{theory}.
The calculations were made for solar metallicity. The
two theoretical models give different predictions, but
the scatter of observed sodium abundances is so
large that it does not allow any conclusions about
the reliability of a particular theoretical model to
be reached. According to the theory, it would be
unreasonable to expect an increase in [Na/Fe] for the
thick-disk stars, because these are low-mass stars.
In contrast, the thin-disk stars have different masses
and, therefore, a large scatter of [Na/Fe] might be
expected for them. To all appearances, this explains
the fact that the thin-disk stars, on average, have an
enhanced [Na/Fe] abundance.

Accordingly, it is hard to say whether the stars
from our sample can be useful in solving the problem
of Galactic chemical evolution. The thick-disk stars,
as being least affected by stellar evolution, will be
more attractive for comparison with the theoretical
models of Galactic evolution.

\section{CONCLUSIONS}

We performed our non-LTE analysis of the sodium
abundance using high-resolution spectra. Based on
our analysis of the (U , V , W ) velocities and the relative abundance [$\alpha$/Fe], we identified four subgroups
of stars: the thin disk (38 stars, 15 of which are Ba~II ones), the thick disk (15 stars), the Hercules
stream (13 stars), and transitional stars (13 stars).
We showed that the departures from LTE should be taken into account even when the weak Na I 6154
and 6161 \,\AA\, lines are used. The non-LTE corrections for Na I 6154 and 6161 \,\AA\, vary between $-$0.06 and
$-$0.24 dex, depending on the stellar parameters. We detected no differences in relative [Na/Fe] abundance between the thick and thin disks and the derived ratios
are nearly solar. The existence of a [Na/Fe] overabundance in Ba~II stars was confirmed. The Hercules-stream stars exhibit nearly solar [Na/Fe] ratios.

\begin{acknowledgments}
This work was financially supported by the Russian Foundation for Basic Research (project no. 14-02-31780 mol-a), the $"$Leading Scientific Schools$"$ Presidential Program (project no. NSh-3620.2014.2),
and the $"$Nonstationary Phenomena in Objects of the Universe$"$ Program of the Presidium of the Russian Academy of Sciences.

\end{acknowledgments}

%\begin{singlespace}

% \end{singlespace}

%\section{перевод специальных терминов на английский язык}

%бариевые звезды -- Ba~II stars
%звезды потока Геркулеса -- Hercules stream stars
%b-фактор  -- departure coefficient
%гиганты сгущения -- clump giants
%термогалинная нестабильность  -- thermohaline instability
%переходые звезды -- transition stars
%химический состав -- chemical abundance
%содержание натрия -- sodium abundance
%тонкая структура -- fine structure
%модель атома -- model atom
%первое перемешивание -- first dredge-up (1DUP)


\begin{thebibliography}{99}

\bibitem{Alves-Brito}
%\refitem{article}
A. Alves-Brito, J. Mel$\acute e$ndez, M. Asplund, I. Ramirez, and D. Yong, Astron. Astrophys. {\bf 513}, A35 (2010).


\bibitem{Andrievsky07}
%\refitem{article}
S. M. Andrievsky, M. Spite, S. A. Korotin, F. Spite, P. Bonifacio, R. Cayrel, V. Hill, and P. Francois, Astron. Astrophys. {\bf 464}, 1081 (2007).


\bibitem{Anstee}
%\refitem{article}
S. D. Anstee and B. J. O\'Mara, Mon. Not. R. Astron.
Soc. {\bf 276}, 859 (1995).


\bibitem{Antipova03}
%\refitem{article}
L. I. Antipova, A. A. Boyarchuk, Yu. V. Pakhomov, et al., Astron. Rep. {\bf 47}, 648 (2003).

\bibitem{Antipova}
%\refitem{article}
L. I. Antipova, A. A. Boyarchuk, Yu. V. Pakhomov, et al., Astron. Rep. {\bf 49}, 535 (2005).

\bibitem{Asplund09}
%\refitem{article}
M. Asplund, N. Grevesse, A. J. Sauval, and P. Scott, Ann. Rev. Astron. Astrophys. {\bf 47}, 481 (2009).


\bibitem{Barklem10}
%\refitem{article}
P. S. Barklem, A. K. Belyaev, A. S. Dickinson, and F. X. Gadea, Astron. Astrophys. {\bf 519}, A20+ (2010).

\bibitem{BarklemOmara}
%\refitem{article}
P. S. Barklem and B. J. O\'Mara, Mon. Not. R. Astron.
Soc. {\bf 290}, 102 (1997).

\bibitem{Butler}
%\refitem{article}
K. Butler and J. Giddings, {Newsletter on the Analysis of Astronomical Spectra}, No. 9 (Univer. London,
1985).

\bibitem{Bau}
%\refitem{article}
D. Baum$\ddot{u}$ller, K. Butler, and T. Gehren, Astron. Astrophys. {\bf 338}, 635 (1998).

\bibitem{Bensby05}
%\refitem{article}
T. Bensby, S. Feltzing, I. Lundstr$\ddot o$m, and I. Ilyin, Astron. Astrophys. {\bf 433}, 185 (2005).

\bibitem{Bensby}
%\refitem{article}
T. Bensby, S. Feltzing, and I. Lundstr$\ddot o$m, Astron.
Astrophys. {\bf 410}, 527 (2003).


\bibitem{Bidelman}
%\refitem{article}
W. P. Bidelman and P. C. Keenan, Astrophys. J. {\bf 114}, 473B (1951).


\bibitem{Boyarchuk1}
%\refitem{article}
A. A. Boyarchuk, L. I. Antipova, M. E. Boyarchuk et
al., Astron. Rep. {\bf 45}, 301 (2001).

\bibitem{BoyarchukLubimkov}
%\refitem{article}
A. A. Boyarchuk and L. S. Lyubimkov, Izv. Krymsk. Astrofiz. Observ. {\bf 64}, 1 (1981).

\bibitem{Bragaglia}
%\refitem{article}
A. Bragaglia, E. Carretta, R. G. Gratton, M. Tosi,
G. Bonanno, P. Bruno, A. Cali, R. Claudi, et al.,
Astrophys. J. {\bf 121}, 327 (2001).


\bibitem{Bruls}
%\refitem{article}
J. H. M. J. Bruls, R. J. Rutten, and N. G. Shchukina,
Astron. Astrophys. {\bf 265}, 237 (1992).

\bibitem{Carretta03}
%\refitem{article}
E. Carretta, A. Bragaglia, C. Cacciari, and E. Rossetti, Astron. Astrophys. {\bf 410}, 143 (2003).

\bibitem{Carretta05}
%\refitem{article}
E. Carretta, A. Bragaglia, R. G. Gratton, and M. Tosi, Astron. Astrophys. {\bf 441}, 131 (2005).


\bibitem{Carretta09}
%\refitem{article}
E. Carretta, A. Bragaglia, R. G. Gratton, and S. Lucatello, Astron. Astrophys. {\bf 505}, 139 (2009).


\bibitem{Cayrel}
%\refitem{article}
R. Cayrel, E. Depagne, M. Spite, V. Hill, F. Spite,
P. Francois, B. Plez, T. Beers, et al., Astron. Astrophys. {\bf 416}, 1117 (2004).

\bibitem{Charbonnel}
%\refitem{article}
C. Charbonnel and N. Lagarde, Astron. Astrophys. {\bf 522}, A10 (2010).


\bibitem{Clayton}
%\refitem{article}
D. D. Clayton, {Handbook of Isotopes in the Cosmos} (Cambridge Univ. Press, Cambridge, 2003).

\bibitem{Cunto}
%\refitem{article}
W. Cunto and C. Mendoza, Rev. Mex. Astron. Astrophis. {\bf 23}, 107 (1992).


\bibitem{Davis}
%\refitem{article}
D. S. Davis, H. B. Richer, I. R. King, J. Anderson,
J. Coffey, G. G. Fahlman, J. Hurley, and J. S. Kalirai,
Mon. Not. R. Astron. Soc. {\bf 383}, L20 (2008).

\bibitem{Dehnen}
%\refitem{article}
W. Dehnen, Astrophys. J. {\bf 115}, 2384 (1998).

\bibitem{Deniss1}
%\refitem{article}
P. A. Denissenkov and S. N. Denisenkova, Astron. Lett. {\bf 16}, 275 (1990).

\bibitem{Deniss2}
%\refitem{article}
P. A. Denissenkov and A. Weiss, Astron. Astrophys.
{\bf 308}, 773 (1996).


\bibitem{Deniss3}
%\refitem{article}
P. A. Denissenkov and F. Herwig, Astrophys. J. {\bf 590},
L99 (2003).

\bibitem{Drawin}
%\refitem{article}
H. W. Drawin, Zeitschr. Phys. {\bf 211}, 404 (1968).

\bibitem{Edvardsson}
%\refitem{article}
B. Edvardsson, J. Andersen, B. Gustafsson,
D. L. Lambert, P. E. Nissen, and J. Tomkin, Astron.
Astrophys. {\bf 275}, 101 (1993).

\bibitem{Famaey}
%\refitem{article}
B. Famaey, A. Jorissen, X. Luri, M. Mayor, S. Udry,
H. Dejonghe, and C. Turon, Astron. Astrophys. {\bf 430},
165 (2005).

\bibitem{Friel}
%\refitem{article}
E. D. Friel, H. R. Jacobson, E. Barrett, L. Fullton,
S. C. Balachandran, and C. A. Pilachowski, Astrophys. J. {\bf 126}, 2372 (2003).

\bibitem{Fuhrmann}
%\refitem{article}
K. Fuhrmann, Astron. Astrophys. {\bf 338}, 161 (1998).

\bibitem{Fux}
%\refitem{article}
R. Fux, Astron. Astrophys. {\bf 373}, 511 (2001).


\bibitem{Gehren}
%\refitem{article}
T. Gehren, Astron. Astrophys. {\bf 38}, 289 (1975).

\bibitem{Girardi}
%\refitem{article}
L. Girardi, A. Bressan, G. Bertelli, and C. Chiosi,
Astron. Astrophys. Suppl. Ser. {\bf 141}, 371 (2000).


\bibitem{Gratton}
%\refitem{article}
R. Gratton, E. Carretta, K. Eriksson, and B. Gustafsson, Astron. Astrophys. {\bf 350}, 955 (1999).

\bibitem{Gratton11}
%\refitem{article}
R. Gratton, S. Lucatello, E. Carretta, A. Bragaglia,
V. D\'Orazi, and Y. Momany, Astron. Astrophys. {\bf 534},
123 (2011).


\bibitem{Gratton12}
%\refitem{article}
R. Gratton, S. Lucatello, E. Carretta, A. Bragaglia,
V. D\'Orazi, Y. Momany, A. Sollima, M. Salaris, and
S. Cassisi, Astron. Astrophys. {\bf 547C}, 2G (2012).

\bibitem{Gustafsson}
%\refitem{article}
B. Gustafsson, B. Edvardsson, K. Eriksson,
U. G. Jorgensen, A. Nordlund, and B. Plez, Astron.
Astrophys. {\bf 486}, 951 (2008).

\bibitem{Iben}
%\refitem{article}
I. Iben, Jr., Astrophys. J. {\bf 147}, 624 (1967).

\bibitem{Igenberg}
%\refitem{article}
K. Igenbergs, J. Schweinzer, I. Bray, et al., At. Data
Nucl. Data Tabl. {\bf 94}, 981 (2008).

\bibitem{Ishigaki}
%\refitem{article}
M. N. Ishigaki, W. Aoki, and M. Chiba, Astrophys.
J. 771, {\bf 771} (2013).

\bibitem{Jacobson1}
%\refitem{article}
H. R. Jacobson, C. A. Pilachowski, and E. D. Friel,
Astrophys. J. {\bf 142}, 59 (2011).

\bibitem{Jacobson07}
%\refitem{article}
H. R. Jacobson, E. D. Friel, and C. A. Pilachowski, Astrophys. J. {\bf 134}, 1216 (2007).

\bibitem{Johnson12}
%\refitem{article}
C. I. Johnson and C. A. Pilachowski, Astrophys. J. {\bf 754}, L38 (2012).

\bibitem{Kacharov}
%\refitem{article}
N. Kacharov, A. Koch, and A. McWilliam, Astron.
Astrophys. {\bf 554A}, 81K (2013).

\bibitem{Kurucz84}
%\refitem{article}
R. L. Kurucz, I. Furenlid, and J. Brault, $Solar Flux
Atlas from 296 to 1300 nm$ (Nat. Solar Observ.,
Sunspot, New Mexico, 1984).

\bibitem{Kurucz93}
%\refitem{article}
R. Kurucz, ATLAS9 Stellar Atmosphere Programs
and 2 km/s grid, Kurucz CD-ROM No. 13 (Smith-
sonian Astrophys. Observ., Cambridge, MA, 1993),
p. 13.

\bibitem{Lind}
%\refitem{article}
K. Lind, M. Asplund, P. S. Barklem, and A. K. Belyaev, Astron. Astrophys. {\bf 528}, A103+
(2011).

\bibitem{Liu}
%\refitem{article}
G. Q. Liu, Y. C. Liang and L. Deng, Res. Astron.
Astrophys. {\bf 9}, 431 (2009).

\bibitem{Luck}
%\refitem{article}
R. E. Luck and U. Heiter, Astrophys. J. {\bf 133}, 2464
(2007).

\bibitem{Lyubimkov}
%\refitem{article}
L. S. Lyubimkov and N. A. Sakhibullin, Astrophysics
{\bf 22}, 203 (1985).

\bibitem{Martin}
%\refitem{article}
W. C. Martin and R. Zalubas, J. Phys. Chem. Ref.
Data {\bf 10}, 153 (1981).

\bibitem{Mashonkina1}
%\refitem{article}
L. I. Mashonkina, N. A. Sakhibullin, and V. V. Shi-
manskii, Astron. Rep. {\bf 37}, 192 (1993).

\bibitem{Mashonkina00}
%\refitem{article}
L. I. Mashonkina, V. V. Shimanskii, N. A. Sakhibullin et al., Astron. Rep. {\bf 44}, 790
(2000).

\bibitem{Mashonkina2}
%\refitem{article}
L. Mashonkina, T. Gehren, J. R. Shi, A. J. Korn, and F. Grupp, Astron. Astrophys. {\bf 528}, A87 (2011).

\bibitem{McClure}
%\refitem{article}
R. D. McClure and A. W. Woodsworth, Astrophys. J. {\bf 352}, 709 (1990).

\bibitem{Mikolaitis}
%\refitem{book}
D. Mihalas, Stellar Atmospheres (Freeman, San Francisco, 1978; Mir, Moscow, 1982), vol. 1, p. 115.

\bibitem{Mikolaitis}
%\refitem{article}
$\check S$. Mikolaitis, G. Tautvai$\check s$iene, R. Gratton, A. Bragaglia, and E. Carretta, Mon. Not. R. Astron. Soc.
{\bf 407}, 1866 (2010).


\bibitem{Mishenina04}
%\refitem{article}
T. V. Mishenina, C. Soubiran, V. V. Kovtyukh, and
S. A. Korotin, Astron. Astrophys. {\bf 418}, 551 (2004).

\bibitem{Mishenina1}
%\refitem{article}
T. V. Mishenina, O. Bienayme, T. I. Gorbaneva,
C. Charbonnel, C. Soubiran, S. A. Korotin, and
V. V. Kovtyukh, Astron. Astrophys. {\bf 456}, 1109 (2006).

\bibitem{Mishenina2}
%\refitem{article}
T. V. Mishenina, C. Soubiran, O. Bienayme, S. A. Korotin, S. I. Belik, I. A. Usenko, and V. V. Kovtyukh,
Astron. Astrophys. {\bf 489}, 923 (2008).

\bibitem{Munoz13}
%\refitem{article}
C. Mu$\tilde n$oz, D. Geisler, and S. Villanova, Mon. Not.
R. Astron. Soc. {\bf 433}, 2006M (2013).

\bibitem{P09}
%\refitem{article}
Yu. V. Pakhomov, L. I. Antipova, A. A. Boyarchuk et
al., Astron. Rep. {\bf 53}, 660 (2009a).

\bibitem{2P09}
%\refitem{article}
Yu. V. Pakhomov, L. I. Antipova, A. A. Boyarchuk et
al., Astron. Rep. {\bf 53}, 685 (2009b).

\bibitem{P11}
%\refitem{article}
Yu. V. Pakhomov, L. I. Antipova, and A. A. Boyarchuk, Astron. Rep. {\bf 55}, 256 (2011).

\bibitem{P12}
%\refitem{article}
Yu. V. Pakhomov, Astron. Lett. {\bf 38}, 101 (2012).

\bibitem{P13}
%\refitem{article}
Yu. V. Pakhomov, Astron. Lett. {\bf 39},
54 (2013).

\bibitem{Park}
%\refitem{article}
C. Park, J. Quant. Spectrosc. Rad. Transfer {\bf 11}, 7 (1971).

\bibitem{Pasquini}
%\refitem{article}
L. Pasquini, S. Randich, M. Zoccali, V. Hill, C. Charbonnel, and B. Nordstrom, Astron. Astrophys. {\bf 424},
951 (2004).

\bibitem{Pereira}
%\refitem{article}
C. B. Pereira, J. V. Sales Silva, C. Chavero, F. Roig, and E. Jilinski, Astron. Astrophys. {\bf 533}, A51 (2011).

\bibitem{Prochaska}
%\refitem{article}
J. X. Prochaska, S. O. Naumov, B. W. Carney, A. McWilliam, and A. M. Wolfe, Astrophys. J. {\bf 120},
2513 (2000).

\bibitem{Reddy}
%\refitem{article}
B. E. Reddy, J. Tomkin, D. L. Lambert, and C. Allende Prieto, Mon. Not. R. Astron. Soc. {\bf 340}, 304 (2003).

\bibitem{Reetz}
%\refitem{article}
J. K. Reetz, Diploma Thesis (Univ. Munchen, 1991).

\bibitem{Rybicki}
%\refitem{article}
G. B. Rybicki and D. G. Hummer, Astron. Astrophys. {\bf 245}, 171 (1991).

\bibitem{Shi}
%\refitem{article}
J. R. Shi, T. Gehren, and G. Zhao, Astron. Astrophys.
{\bf 423}, 683 (2004).

\bibitem{Smiljanic}
%\refitem{article}
R. Smiljanic, R. Gauderon, P. North, B. Barbuy, C. Charbonnel, and N. Mowlavi, Astron. Astrophys.
{\bf 502}, 267 (2009).

\bibitem{Smiljanic12}
%\refitem{article}
R. Smiljanic, Mon. Not. R. Astron. Soc. {\bf 422}, 1562S
(2012).

\bibitem{Soubiran}
%\refitem{article}
C. Soubiran and P. Girard, Astron. Astrophys. {\bf 438},
139 (2005).

\bibitem{Takeda}
%\refitem{article}
Y. Takeda, B. Sato, and D. Murata, Publ. Astron. Soc. Jpn. {\bf 60}, 781 (2008).


\bibitem{Takeda03}
%\refitem{article}
Y. Takeda, G. Zhao, M. Takada-Hidai, Yu-Qin Chen,
Yu-Ji Saito, and Hua-Wei Zhang, Chin. J. Astron.
Astrophys. {\bf 3}, 316 (2003).


\bibitem{Woosley}
%\refitem{article}
S. E. Woosley and T. A. Weaver, Astrophys. J. Suppl.
Ser. {\bf 101}, 181 (1995).

\bibitem{Yong}
%\refitem{article}
D. Yong, F. Grundahl, P. E. Nissen, H. R. Jensen, and
D. L. Lambert, Astron. Astrophys. {\bf 438}, 875 (2005).

\bibitem{Zwicky}
%\refitem{book}
F. Zwicky, Morphological Astronomy (Springer,
1957), p. 258.



\end{thebibliography}
\end{document}